\documentclass[aps,superscriptaddress,eqsecnum,nofootinbib,showpacs,preprintnumbers]
{revtex4}
\usepackage{graphicx,epsfig}
\usepackage{amsmath}
\usepackage {amssymb}

\newcommand{\be}{\begin{eqnarray}}
\newcommand{\ee}{\end{eqnarray}}
\newcommand{\bea}{\begin{eqnarray}}

\newcommand{\eea}{\end{eqnarray}}

\makeindex

\begin{document}


\title{Superradiant Instability of Near Extremal and Extremal Four-Dimensional  Charged Hairy   Black Hole in anti-de Sitter Spacetime}
\author{P. A. Gonz\'{a}lez}
\email{pablo.gonzalez@udp.cl} \affiliation{Facultad de
Ingenier\'{i}a y Ciencias, Universidad Diego Portales, Avenida Ej\'{e}rcito
Libertador 441, Casilla 298-V, Santiago, Chile.}
\author{Eleftherios Papantonopoulos}
\email{lpapa@central.ntua.gr} \affiliation{Department of
Physics, National Technical University of Athens, Zografou Campus
GR 157 73, Athens, Greece.}
\author{Joel Saavedra}
\email{joel.saavedra@ucv.cl} \affiliation{Instituto de
F\'{i}sica, Pontificia Universidad Cat\'olica de Valpara\'{i}so,
Casilla 4950, Valpara\'{i}so, Chile.}
\author{Yerko V\'{a}squez}
\email{yvasquez@userena.cl}
\affiliation{Departamento de F\'{\i}sica, Facultad de Ciencias, Universidad de La Serena,\\
Avenida Cisternas 1200, La Serena, Chile.}
\date{\today}

\vspace{0.5cm}

\begin{abstract}

We study the instability of near extremal and extremal four-dimensional  AdS charged hairy black hole to radial  neutral massive and charged massless scalar field perturbations. We solve the scalar field equation by using the improved asymptotic iteration method and the time domain analysis and we find the quasinormal frequencies. For the charged scalar perturbations, we find the superradiance condition by computing the reflection coefficient in the low-frequency limit and we show that in the  superradiance regime, which depends on the scalar hair charge,  all modes of radial charged massless perturbations are unstable, indicating  that the charged hairy black hole is superradiantly unstable. On the other hand, calculating the quasinormal frequencies of radial neutral scalar perturbations in this background, we found stability of the charged hairy black hole.

\end{abstract}

\maketitle


\tableofcontents


\section{Introduction}

In Gravity theories the stability of black holes has been a central issue and it has been studied for a long time starting from the pioneering work
by Regge and Wheeler \cite{Regge:1957a}. It has been found that most of black holes are stable under  various types of
perturbations.  An effective way to study the stability of black holes is to calculate the  quasinormal modes (QNMs) and their quasinormal frequencies (QNFs) \cite{Zerilli:1971wd,
Zerilli:1970se, Kokkotas:1999bd, Nollert:1999ji}. The QNMs and QNFs give
information about the stability of matter fields that
evolve perturbatively in the exterior region of a black hole without backreacting on the
metric but they carry the information of the black hole because they  depend only on their parameters  (mass, charge and angular momentum) and the fundamental constants (Newton constant and cosmological constant) of their spacetime, just like the parameters that define the test field.
 The QNFs have been calculated by means of numerical and analytical techniques. Some well known numerical methods are: the Mashhoon method, Chandrasekhar-Detweiler method, WKB method, Frobenius method, method of continued fractions, Nollert asymptotic iteration method (AIM) and improved AIM, among others (for reviews on   QNMs and their QNFs see \cite{Berti:2004md,Konoplya:2011qq}).

 Extensive study of QNMs of black holes in
asymptotically flat spacetimes have been performed for the last few
decades mainly due to the potential
astrophysical interest. Considering the case when the black hole is
immersed in an expanding universe, the QNMs of black holes in de
Sitter space have also been investigated \cite{deSitter_1,deSitter_2}. The holographic understanding of various system using the
AdS/CFT correspondence has triggered recently the study of QNMs in
anti-de Sitter (AdS) spacetimes.  The first study of the
QNMs in AdS spaces was performed in
\cite{Chan_Mann}. Subsequently, in \cite{Horowitz:1999jd} it was suggested a
numerical method to calculate the QNFs directly and made a
systematic investigation of QNMs for scalar perturbation on the
background of Schwarzschild-AdS  black holes. Later the numerical method of \cite{Horowitz:1999jd}
 was generalized to the study of QNMs of  Reissner-Nordstr\"om-AdS black holes in \cite{Wang-00}.
Also this method was also generalized and extended using the time domain analysis approach
\cite{Wang-01}.

The application of gauge/gravity duality  to condense matter systems  put forward the necessity to study the behaviour of scalar fields just outside the black hole horizon \cite{Gubser:2005ih,Gubser:2008px}. The simplest understanding of this application is the holographic superconductor \cite{Hartnoll:2008vx}. According to the duality principle, a scalar field minimally coupled to gravity,  below a critical temperature condenses outside the black hole horizon, destabilizing the black hole spacetime, and the black hole acquires scalar hair. This mechanism also works if the scalar field is not minimally  coupled to gravity \cite{Kuang:2016edj}. Another interesting application of the  gauge/gravity duality is to look on  condensed matter systems at criticality
where quantum properties of these systems at zero temperature can be studied \cite{Horowitz:2009ij}. The gravity part of these
systems is  described by an extremal charged black hole in anti-de
Sitter spacetime \cite{Chamblin:1999tk}.  The metric of the
extremal black hole has  interesting properties. In the near
horizon limit when the temperature goes to zero the horizon
geometry is given by AdS$_2 \times R^{2}$. In
connection to holographic superconductors the near horizon
geometry of an extremal black hole was studied in \cite{Kunduri:2013gce}.

The above discussion suggests that it is important to study the stability of charged AdS black holes in the presence of scalar hair  as the temperature approaches zero.  This requires the calculation of  the QNMs and QNFs of near extremal and extremal black holes in AdS spacetimes. However, the method developed in \cite{Horowitz:1999jd} breaks down  for large
values of the charge of the Reissner-Nordstr\"om-AdS black holes \cite{Berti-03,Wang-00,Wang-01, Mahapatra:2016dae}. Therefore, for large black holes to calculate  the QNMs and QNFs of  extremal black holes we have to use other methods, like the time  domain analysis.

Going to zero temperature at the extremal limit we have also to deal with the applicability of the known methods. In most of the methods calculating the QNMs, a perturbative expansion around the even horizon of the black hole was used. However, the event horizon of an extremal black hole is an irregular singular point of the perturbation equation. Therefore, unlike the case of non-extremal black holes, one does not expect a non-zero radius of convergence of the associated power series around the event horizon. A solution to this problem was proposed in \cite{Onozawa:1995vu} in which an expansion was carried out  around an ordinary point of the differential equation. More recently in \cite{Richartz:2015saa}, the modified version of the continued fraction method of  \cite{Onozawa:1995vu} was used and the QNMs of charged massless perturbations around an extremal Reissner-Nordstr\"om black hole and of neutral massless perturbations around an extremal Kerr black hole were calculated.

The behaviour of matter fields near  the  horizon
of a charged black hole at the zero temperature limit was studied in \cite{Gonzalez:2014tga}. To evade the  no-hair theorems and have  a healthy
behaviour of the scalar field, regular on the horizon and  fall
off sufficiently fast at large distances, a profile for the
scalar field was introduced and a fully backreacted solution of the
Einstein-Maxwell-scalar system was found.   Also, in \cite{O.:2016wcf} charged hairy black holes in non-linear electrodynamics were found. Exact solutions of this system without the
electromagnetic field were found in \cite{Gonzalez:2013aca}. Hairy black holes and their stability were also studied  in \cite{Dias:2016pma}.

In this work, we study the instability of a near extremal and extremal four-dimensional charged hairy  black hole in AdS spacetime using the solution found in \cite{Gonzalez:2014tga}. In this background we consider a scalar wave minimally coupled to gravity. Then, studying the radial  neutral massive and charged massless scalar wave perturbations, we  obtain numerically the QNFs of these perturbations. To solve the scalar field equation we use the improved AIM \cite{Cho:2009cj}, which is an improved version of the method proposed in  \cite{Ciftci,Ciftci:2005xn} and it has been applied successfully in the context of QNMs for different black hole geometries (see for instance \cite{Koutsoumbas:2006xj, Cho:2009cj, Cho:2011sf, Catalan:2013eza, Catalan:2014ama, Zhang:2015jda, Barakat:2006ki, Sybesma:2015oha, Gonzalez:2015gla, Becar:2015gca, Becar:2015kpa}). To obtain the QNFs of the extremal hairy black hole we use the time domain analysis.

 For charged scalar perturbations we calculate the superradiance condition by computing the reflection coefficient in the low-frequency limit. We show that the Bekenstein superradiance condition \cite{Bekenstein}  is modified incorporating the information of the presence of scalar hair in the theory. We find that there is a critical value $q_c$ of the charge of the scalar field above which all the modes of radial charged massless perturbations are unstable, which is related to the charge of the scalar hair, and we show that in this regime the superradiance condition is satisfied, therefore   the charged hairy black hole is superradiantly unstable. By calculating the QNFs of radial neutral massless scalar perturbations we find stability, as expected.

The work is organized as follows. In Section~\ref{Background} we give a brief review
 of the four-dimensional charged hairy  black hole found in \cite{Gonzalez:2014tga}.
 In Section~\ref{scalar pert}
we calculate the QNFs of  scalar  perturbations numerically, by using the improved AIM method in Section~\ref{AIM} and the time domain analysis in Section~\ref{TDA}. In Section~\ref{superrandiant} we study the superradiant effect calculating the superradiance condition. Finally, our conclusions are in Section~\ref{conclusion}.

\section{Four-Dimensional Charged Hairy Black Hole}
\label{Background}

 In this section we will review the charged hairy black hole solution discussed in \cite{Gonzalez:2014tga}. This is a solution of a theory that consists of a scalar field minimally coupled to curvature having a self-interacting potential $ V(\phi)$, in the
presence of an electromagnetic field.
 The Einstein-Hilbert action
with  a negative cosmological constant $\Lambda=-6l^{-2}/\kappa$,
where $l$ is the length of the AdS space, which it has been incorporated in
the potential as $ \Lambda=V(0)$ ($V(0)<0$), reads
 \begin{eqnarray} \label{action}
 S=\int d^{4}x\sqrt{-g}\left(\frac{1}{2 \kappa }R-\frac{1}{4}F_{\mu \nu}F^{\mu \nu}
 -\frac{1}{2}g^{\mu\nu}\nabla_{\mu}\phi\nabla_{\nu}\phi-V(\phi)\right)~,
 \end{eqnarray}
where $\kappa=8 \pi G_N$, with $G_N$ the Newton constant. The
resulting Einstein equations from the above action are
 \begin{eqnarray}
 R_{\mu\nu}-\frac{1}{2}g_{\mu\nu}R=\kappa (T^{(\phi)}_{\mu\nu}+T^{(F)}_{\mu\nu})~,\label{field1}
 \end{eqnarray}
where the energy momentum tensors $T^{(\phi)}_{\mu\nu}$ and $T^{(F)}_{\mu\nu}$ for the
scalar and electromagnetic fields are
 \begin{eqnarray}
 \nonumber T^{(\phi)}_{\mu\nu}&=&\nabla_{\mu}\phi\nabla_{\nu}\phi-
 g_{\mu\nu}[\frac{1}{2}g^{\rho\sigma}\nabla_{\rho}\phi\nabla_{\sigma}\phi+V(\phi)]~,\\
  T^{(F)}_{\mu\nu}&=&F_{\mu}^{\alpha}F_{\nu \alpha} -\frac{1}{4}g_{\mu \nu}F_{\alpha\beta}F^{\alpha\beta}  \label{energymomentum}~.
 \end{eqnarray}
 If we use Eqs. (\ref{field1}) and (\ref{energymomentum}) we obtain the equivalent equation
 \begin{eqnarray}
 R_{\mu\nu}-\kappa\left(\partial_\mu\phi \partial_\nu\phi+g_{\mu\nu}V(\phi)\right)=\kappa (F_{\mu}^{\alpha}F_{\nu \alpha} -\frac{1}{4}g_{\mu \nu}F_{\alpha\beta}F^{\alpha\beta})~. \label{einstein1}
 \end{eqnarray}
We consider the following metric ansatz
 \begin{eqnarray}
 ds^{2}=-f(r)dt^{2}+f^{-1}(r)dr^{2}+a^{2}(r)d \sigma^2 ~,\label{metricBH}
 \end{eqnarray}
where $d \sigma ^2$ is the metric of the spatial 2-section, which
can have positive, negative or zero  curvature, and
$A_{\mu}=(A_t(r),0,0,0)$ the scalar potential of the
electromagnetic field.

 A particular
profile of the scalar field was considered in \cite{Gonzalez:2014tga}
\begin{equation}
\phi \left( r\right) =\frac{1}{\sqrt{2}}\ln \left( 1+\frac{\nu }{r}\right) ~,
\label{field}
\end{equation}
where $\nu $ is the scalar charge, a parameter controlling the behaviour of the
scalar field and it has the dimension of length and then the following analytical solution was found
\begin{eqnarray}\label{ametric}
\nonumber a\left( r\right)& =&\sqrt{r\left( r+\nu \right) }~,\\
A_t(r)&=&\frac{Q}{\nu}ln\left(\frac{r}{r+\nu}\right)~.
\end{eqnarray}
The metric function $f(r)$ is
\begin{eqnarray}
 f\left( r\right) &= &-2\frac{Q^2}{\nu^2}+C_1r(r+\nu)-
\frac{C_2(2r+\nu)}{\nu^2}+2\frac{k r(2r+\nu)} {\nu^2} \nonumber
\\
&& -2\left(\frac{ Q^2(2r+\nu)+r(r+\nu)(C_2+k\nu)
}{\nu^3}+\frac{Q^2r(r+\nu)ln\frac{r}{r+\nu}}{\nu^4}\right)
ln\frac{r}{r+\nu}~, \label{f(r)}
\end{eqnarray}
where $k=-1,0,1$  parameterizes the curvature of the transverse 2-section and $C_1$, $C_2$ are  integration constants being proportional to
the cosmological constant and to the mass respectively,
and the potential is given by
\begin{eqnarray}
\nonumber V\left( \phi \right) &=&  \frac{1}{2\nu^4}e^{-2\sqrt{2}\phi}\left(\left(e^{-2\sqrt{2}\phi}-1\right)^2\left(1+10e^{\sqrt{2}\phi} +e^{\sqrt{2}\phi}\right)Q^2 \right. \nonumber
\\
&&
+e^{\sqrt{2}\phi}\nu \left(-6C_2-10k\nu-C_1\nu^3-4e^{\sqrt{2}\phi}\nu(4k+C_1\nu^2)+e^{2\sqrt{2}\phi}(6C_2+2k\nu-C_1\nu^3)\right)\nonumber
\\
&&
+2coth\frac{\phi}{\sqrt{2}}\left(\left(1+4e^{\sqrt{2}\phi}+e^{2\sqrt{2}\phi}\right)Q^2ln\frac{\nu}{-1+e^{\sqrt{2}\phi}} \right.\nonumber
\\
&&
\left. 2e^{\sqrt{2}\phi} \left( \left(2+cosh\sqrt{2}\phi \right)\left( \nu(C_2+k\nu)+Q^2ln(1-e^{\sqrt{2}\phi})+6Q^2sinh\sqrt{2}\phi\right)\right) \right),
\label{m2potn}
\end{eqnarray}
where $ V\left( 0\right) =\Lambda_{eff}$  as expected and also
\begin{equation} C_1+\frac{4k}{\nu^2}=-\frac{\Lambda_{eff}
}{3}=\frac{1}{l^{2}}~. \label{relation}\end{equation} We see from
the above relation that the parameter $\nu$ of the scalar field
introduces a length scale connected with the presence of the
scalar field in the theory.
Besides, we know that%
\begin{equation}
V^{\prime \prime }\left( \phi =0\right) =m^{2}~,
\end{equation}
where $m$ is the scalar field mass. Therefore, we obtain that the
scalar field mass is given by
\begin{equation}
m^{2}=\frac{2}{3}\Lambda_{eff} =-2l^{-2}~,
\end{equation}
which satisfies the Breitenlohner-Friedman bound that ensures the
perturbative stability of the AdS spacetime
\cite{Breitenlohner:1982jf}.

One may wander if in the limit of $\Lambda_{eff} \rightarrow 0$
and $\nu \rightarrow 0$ we recover the Reissner-Nordstr\"om (RN)
black hole. Indeed from  (\ref{f(r)}) if we fix the constant $C_1$
to $C_1=-\frac{4k}{\nu^2}$  the function $f(r)$ can be written as
\begin{eqnarray}
f(r)&=& -\frac{2\left(Q^2+5\nu+r(C_2+k\nu)\right)}{\nu^2} \nonumber
\\
&&-\frac{2\left(\nu\left(Q^2(2r+\nu)+r(r+\nu)(C_2+k\nu)\right)+Q^2r(r+\nu)ln\frac{r}{r+\nu}\right)ln\frac{r}{r+\nu}}{\nu^4}~,
\label{RNL1}
\end{eqnarray}
and in the limit  $\nu \rightarrow 0$ we recover the RN black hole
\begin{equation}\label{RNL2}
f(r)=k+\frac{Q^2}{2r^2}-\frac{C_2}{3r}~.
\end{equation}

It was shown, that in the case of $\Lambda_{eff}=  0$ at all temperature the RN black hole solution is thermodynamically preferred over the charged hairy  black hole solution. In the case
of $\Lambda_{eff} \neq 0$ the charged hairy  black
hole is thermodynamically preferred over the RN black hole at low
temperature. This picture is in agreement with the findings of the
application of the AdS/CFT correspondence to condensed matter
systems. In these systems there is a critical temperature below
which the system undergoes a phase transition to a hairy black
hole configuration at low temperature. This corresponds in the
boundary field theory to the formation of a condensation of the
scalar field  \cite{Hartnoll:2008vx,Koutsoumbas:2009pa}. On the other hand, classically, it was shown that  there are bounded orbits like planetary orbits in this background. Also, the periods associated to circular orbits are modified by the presence of the scalar hair. Besides, some classical tests such as perihelion precession, deflection of light and gravitational time delay have the standard value of general relativity plus a correction term coming from the cosmological constant and the scalar hair. Furthermore, it is possible to find a specific value of the parameter associated to the scalar hair, in order to explain the discrepancy between the theory and the observations, for the perihelion precession of Mercury and light deflection \cite{Gonzalez:2015jna}.

Now, we make the following redefinitions
\begin{eqnarray}
\nonumber C_1&=&-\frac{\Lambda}{3}-\frac{4k}{\nu^2}~,\\
\nonumber C_2&=&\alpha_2\nu^3-k\nu~,\\
Q^2&=&\alpha_1\nu^4~. \label{redef}
\end{eqnarray}
 Then the potential becomes
\bea
 V(\phi) &=& \alpha_1 (8 \phi^2+4 \left(\phi^2+2\right) \cosh \left(\sqrt{2} \phi \right)-12 \sqrt{2} \phi  \sinh \left(\sqrt{2} \phi \right)
  +\cosh \left(2 \sqrt{2} \phi \right)-9 )\nonumber \\
&&+6\alpha_2 \sinh \left(\sqrt{2} \phi \right)-2 \sqrt{2} \alpha_2 \phi  \left(\cosh \left(\sqrt{2} \phi \right)+2\right)+\frac{1}{3} \Lambda
 \left(\cosh \left(\sqrt{2} \phi \right)+2\right) \,,
\eea
while the metric function is
\bea
 f(r)&=&-\frac{\Lambda r^2}{3}-\left(\frac{\Lambda}{3}+2\alpha_2 \right)\nu r - (2\alpha_1+\alpha_2)\nu^2+k \nonumber \\
&&-2(\alpha_1 \nu^2 +\alpha_2 r^2 +\nu r (2\alpha_1+\alpha_2))\ln(\frac{r}{r+\nu})-2\alpha_1 r (r+\nu)\ln^2(\frac{r}{r+\nu})\,. \label{redfun}
\eea
The scalar field is
\begin{equation}
\phi=\frac{1}{\sqrt{2}}\ln(1+\frac{\nu}{r})~,
\end{equation}
and the electric potential becomes
\begin{equation}
A_t=\sqrt{\alpha_1}\nu \ln (\frac{r}{r+\nu})~.
\end{equation}

With this redefinition  the solution has one integration constant $\nu$ (the electric charge is not independent) and the self interaction potential has three parameters, the cosmological constant $\Lambda$ and two coupling constants $\alpha_1$ and $\alpha_2$.

\section{Scalar perturbations }
\label{scalar pert}

\begin{figure}[h]
\begin{center}
\includegraphics[width=0.7\textwidth]{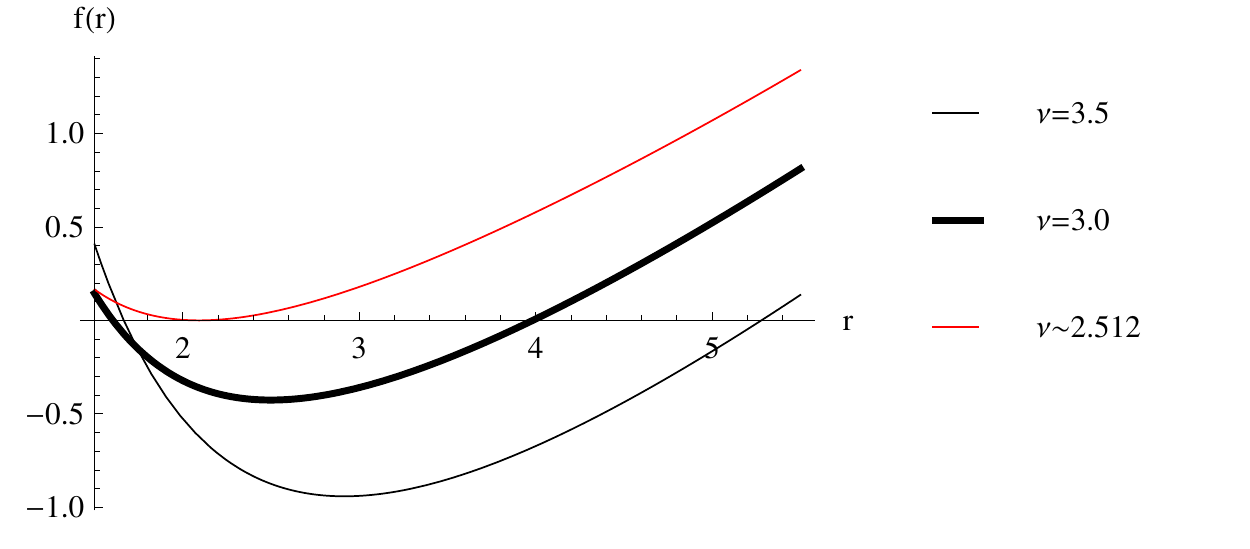}
\end{center}
\caption{The behavior of $f(r)$ with $\alpha_1= 1$, $\alpha_2=2$, $m=0.1$, $q=0$, $k=1$, $\kappa=0$, and $\Lambda=-0.1$.} \label{Function1}
\end{figure}

\begin{figure}[h]
\begin{center}
\includegraphics[width=0.7\textwidth]{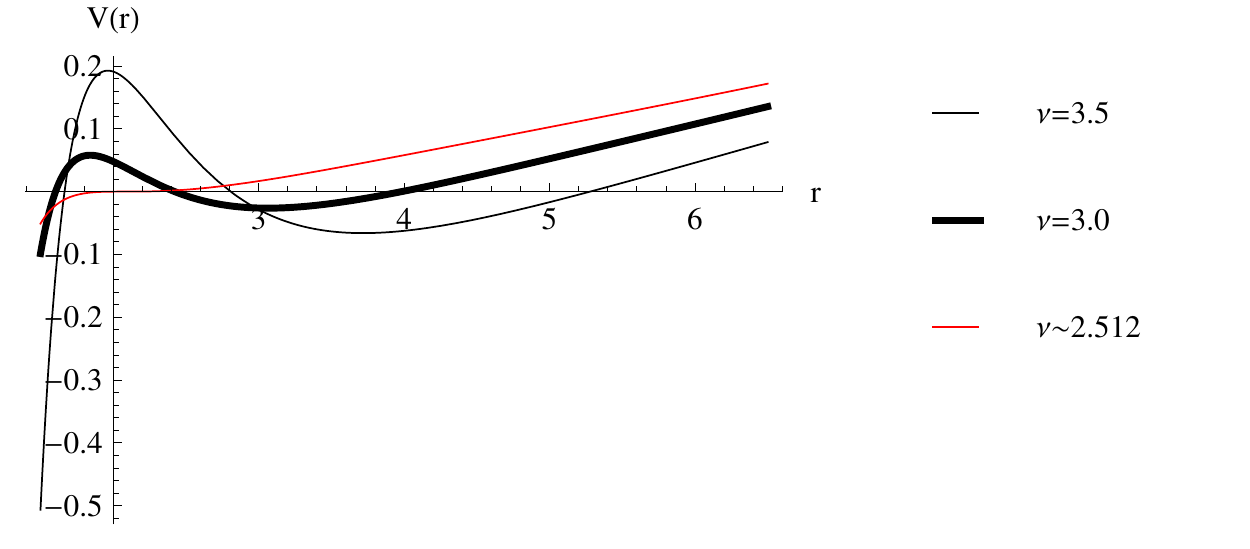}
\end{center}
\caption{The behavior of $V_{eff}(r)$ with $\alpha_1=1$, $\alpha_2=2$, $m=0.1$, $q=0$, $k=1$, $\kappa=0$, and $\Lambda=-0.1$.} \label{Potential11}
\end{figure}

The QNMs of charged scalar perturbations in the background of  the metric (\ref{redfun})
are given by the scalar field solution of the Klein-Gordon equation
\begin{equation}
\frac{1}{\sqrt{-g}} \left(\partial _{\mu }-iqA_{\mu} \right) \left( \sqrt{-g}g^{\mu \nu } \left(\partial
_{\nu }-iqA_{\nu} \right) \varphi \right) =m^{2}\varphi \,,  \label{KGNM}
\end{equation}
with suitable boundary conditions for a black hole geometry. In the above expression $m$ is the mass and $q$ is the electric charge of the scalar field $\varphi $. In the following we will focus our attention to spherical transverse section ($k=1$). Now, by means of the following ansatz
\begin{equation}
\varphi =e^{-i\omega t} R(r) Y(\Omega) \,,\label{wave}
\end{equation}%
the Klein-Gordon equation reduces to
\begin{equation}
\frac{d}{dr}\left(a(r)^2 f(r)\frac{dR}{dr}\right)+\left(\frac{a(r)^2 \left( \omega+q A_{t}(r) \right)^2}{f(r)}-\kappa^2-m^{2}a(r)^2 \right) R(r)=0\,, \label{radial}
\end{equation}%
where we have defined $-\kappa^2=-\ell (\ell+1)$, with $\ell=0,1,2,...$, which represents the eigenvalue of the Laplacian on the two-sphere.
Now, defining the radial function $R(r)$ as
 \begin{equation}
 R(r)=\frac{F(r)}{a(r)}\,,
 \end{equation}
and by using the tortoise coordinate $r^*$ given by
 \begin{equation}\label{ast}
 dr^*=\frac{dr}{f(r)}\,,
 \end{equation}
 the Klein-Gordon equation can be written as a one-dimensional Schr\"{o}dinger equation
 \begin{equation}\label{ggg}
 \frac{d^{2}F(r^*)}{dr^{*2}}-V_{eff}(r)F(r^*)=-\omega^{2}F(r^*)\,,
 \end{equation}
 with an effective potential $V_{eff}(r)$, which is parametrically written as $V_{eff}(r^*)$ and it is given by
  \begin{equation}\label{pot}
 V_{eff}(r)=\frac{f(r)}{a(r)^2} \left(\kappa^2 + a(r)\left( m^2a(r)+  a^\prime(r)f^\prime(r)+a^{\prime\prime}(r)f(r)\right) \right)-2 q \omega A_t(r) - q^2 A_t (r) ^2~.
 \end{equation}
For neutral scalar field $q=0$, the effective potential is real and diverges at spatial infinity. In Fig. \ref{Function1} we plot the lapsus function and in Fig. \ref{Potential11} the potential, where we have considered a spherical transverse section ($k=1$), massive scalar fields with $m=0.1$, $\alpha_1=1$, $\alpha_2=2$ $\Lambda=-0.1$, $\kappa=0$, and $\nu =3.5, 3.0, 2.512$, for $\nu \approx 2.5119365 \approx 2.512$ the black hole is extremal and then we choose different values of $\nu$ in order to consider near extremal black holes. The event horizon for the extremal black hole is localized at $r_{+} \approx 2.0921064 \approx 2.09$.


In order to know about the stability of scalar fields in the background of a black hole, we follow a  general argument given in Ref. \cite{Horowitz:1999jd}.
So, by defining
\begin{equation}
\psi(r)=e^{i\omega r^{\ast}} F(r)\,,
\end{equation}
and inserting this expression in the Schrodinger like equation \eqref{ggg} yields
\begin{equation}\label{KleinFink}
\frac{d}{dr}(f(r)\frac{d\psi(r)}{dr})-2i\omega \frac{d\psi(r)}{dr}-\frac{V_{eff}(r)}{f(r)}\psi(r)=0\,.
\end{equation}
Then, multiplying Eq. (\ref{KleinFink}) by $\psi^{\ast}$ and performing integrations by parts, and  additionally using the Dirichlet boundary condition for the scalar field at spatial infinity, one can obtain the following expression
\begin{equation}\label{relacion}
\int _{r_{+}}^{\infty}dr \left( f(r) \left|  \frac{d\psi}{dr}\right|^2+\frac{V_{eff}(r) |_{q=0}}{f(r)} \left| \psi \right| ^2 -\frac{q^2 A_{t}(r)^2}{f(r)} \left| \psi \right| ^2  \right)=-\frac{\left|\omega \right|^2 \left| \psi (r=r_{h})\right| ^2}{Im(\omega)}\,.
\end{equation}
In general, the QNFs are complex, where the real
part represents the oscillation frequency and the imaginary part describes the rate at which this oscillation is damped, with the stability of the scalar field being guaranteed if the imaginary part is negative.
Notice that the sign outside the horizon of the expression $V_{eff}|_{q=0}-q^2 A_{t}^2$, which appears inside the integral of the above equation, is crucial for stability. For the neutral scalar perturbations the effective potential (\ref{pot})
 is positive outside the horizon and
then the left hand side of \eqref{relacion} is strictly positive, which demand that $Im (\omega)<0$, and then we conclude that the stability of the neutral scalar field under perturbations respecting Dirichlet  boundary conditions is obeyed. For charged scalar field, the integral can yield a negative value, therefore the stability is not guaranteed in this case.


In the following  we will compute numerically the QNFs by using the improved AIM for hairy black hole. However, as we discussed in the introduction the improved AIM breaks down  when we approach  the extremal limit.  Therefore, we will use the time domain analysis to compute numerically the QNFs in this limit.


\subsubsection{Calculation of QNFs using the improved AIM}
\label{AIM}





Now, in order to implement the improved AIM (we give a brief review of the improved AIM in the Appendix)  we perform the  change of variable
$y=1-\frac{r_{+}}{r}$. So, the Klein-Gordon equation reads
\begin{equation}
R^{\prime \prime}(y) +\left( 2 \frac{a^{\prime}(y)}{a(y)}+\frac{f^{\prime} (y)}{f(y)}-\frac{2}{1-y} \right) R^{\prime} (y)+  \frac{r_{+}^2}{(1-y)^4} \left(\frac{\left( \omega + q A_t(y) \right)^2}{f(y)^2} -\frac{\kappa^2}{a(y)^2 f(y)} -\frac{m^2}{f(y)} \right) R(y) =0 \, .
\label{numericalmethod}
\end{equation}
In this equation the functions $f(y)$, $a(y)$ and $A_t (y)$ refers to the functions $f(r)$, $a(r)$ and $A_t (r)$ evaluated at $r=\frac{r_{+}}{1-y}$, and
we will study the behaviour of the above equation at the horizon and at spatial infinite.
At the event horizon $y=0$, the Eq. (\ref{numericalmethod}) can be written as
\begin{equation}
R ^{\prime \prime}(y)+\frac{1}{y} R^{\prime}(y)+\frac{ r_+^2\left( \omega +qA_t (0)\right) ^2}{f^\prime(0)^2 y^2}R(y)=0\,,
\end{equation}
where we have considered the Taylor expansion of the lapse function around the event horizon $f(y) \approx f^\prime(0)y+ \mathcal{O}(y^2)$.
 Then, the  solution is given by
\begin{equation}
R(y)=C_1e^{\frac{i r_+  (\omega+qA_t (0)) \ln y}{f^\prime(0)}}+C_2e^{-\frac{i r_+  (\omega+qA_t (0)) \ln y}{f^\prime(0)}}\,,
\end{equation}
which it can be written in terms of the tortoise coordinate $r^{\ast}$, and near the horizon is given by $r^{\ast}=\frac{r_+ }{f^\prime(0)} \ln y$, as
\begin{equation}\label{eq1}
R(y)=C_{1}e^{i (\omega+q A_t (0)) r^{\ast}}+C_{2}e^{-i(\omega+qA_t (0)) r^{\ast}}\,.
\end{equation}
So, by imposing as a boundary condition that only ingoing waves exist on the event horizon, we must set $C_{1}=0$. Therefore, the solution near the horizon is given by
\begin{equation}\label{horizon}
R(y)=C_2e^{-\frac{i r_+  (\omega+qA_t (0)) \ln y}{f^\prime(0)}}= C_2 y^{-\frac{i r_+ (\omega +q A_t (0))}{f^{\prime} (0)}}  \,.
\end{equation}
On the other hand, at spatial infinity, the Klein-Gordon equation becomes
\begin{equation}
R^{\prime \prime} (y)+\frac{2}{1-y}R^{\prime}(y)+\frac{3 m^2}{\Lambda (1-y)^2}R(y)=0\,,
\end{equation}
whose solution is
\begin{equation}
R(y)=D_1(1-y)^{\beta_+}+D_2r (1-y) ^{\beta_-}\,,
\end{equation}
where $\beta_{\pm}=\frac{3}{2}\pm\sqrt{\left(  \frac{3}{2}\right)^2-\frac{3m^2}{\Lambda}}$. So, imposing the Dirichlet boundary condition; that is, having a null scalar field at spatial infinity, we must set $D_{2}=0$.
Therefore, in order to implement the above boundary conditions in the improved AIM, we must redefine the radial function $R(y)$ in terms of a new function, say $\chi (y)$, in the following form
\begin{equation}
R\left( y\right) = y^{\alpha} (1-y)^{\beta} \chi (y)\,,
\end{equation}
where
\begin{eqnarray}
\nonumber \alpha&=&-\frac{i r_+ (\omega+qA_t (0))}{f'(0)}~, \\
\nonumber \beta&=&\frac{3}{2}+\sqrt{\left( \frac{3}{2} \right)  ^2-\frac{3m^2}{\Lambda}} \, .
\end{eqnarray}

Then, by inserting the above expression for $R(y)$ in Eq. (\ref{numericalmethod}), we obtain the following homogeneous linear second-order differential equation for the function $\chi (y)$
\begin{equation}
\chi ^{\prime \prime }(y)=\lambda _{0}(y)\chi ^{\prime }(y)+s_{0}(y)\chi(y) \,,
\label{de}
\end{equation}
where
\begin{equation}
\lambda _{0}(y)=-\frac{2 f(y) \left((y-1) y a'(y)+a(y) (y (\alpha +\beta +1)-\alpha )\right)+(y-1) y a(y) f'(y)}{(y-1) y a(y) f(y)}\,,
\end{equation}
\begin{eqnarray}
\notag  s_{0}(y)&=& - \frac{\beta  (\beta +1) y^2+\alpha ^2 (y-1)^2+\alpha  (y-1) (2 \beta  y+y+1)}{(y-1)^2 y^2}\\
&& - \frac{2 (y-1)^3 a(y) f(y) a'(y) (\alpha  (y-1)+\beta  y)-\kappa^2  r_+ ^2 y}{(y-1)^4 y a(y)^2 f(y)}  \\
&&  - \frac{r_+ ^2 y \left((\omega+q A_t (y))^2-m^2 f(y)\right)+(y-1)^3 f(y) f'(y) (\alpha  (y-1)+\beta  y)}{(y-1)^4 y f(y)^2} \,.
\end{eqnarray}

\begin{figure}[h]
\begin{center}
\includegraphics[width=0.45\textwidth]{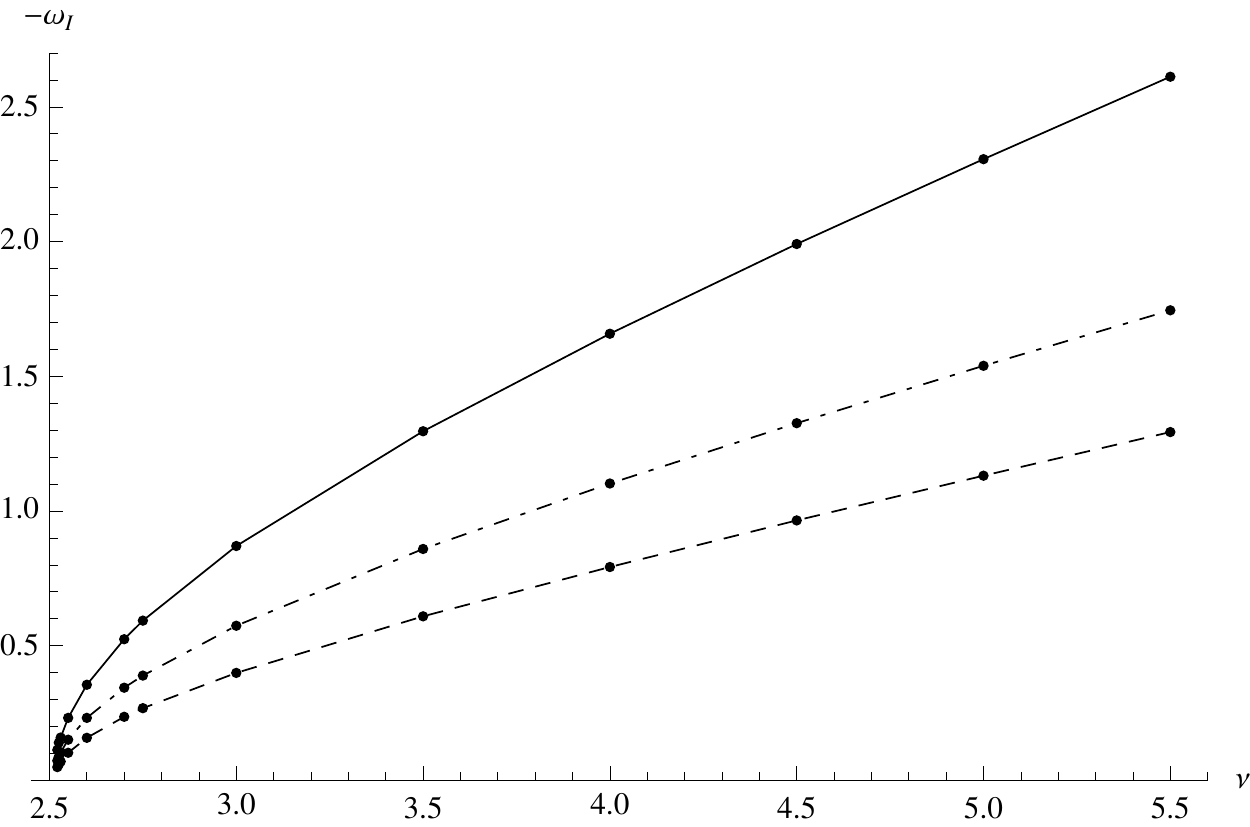}
\end{center}
\caption{The behaviour of $-\omega_{I}$ for $\kappa=0$, $\alpha_1=1$, $\alpha_2=2$, $\Lambda=-0.1$, $q=0$, $
k=1$ and different values of $m$ and $\nu$. $m=0$ (dashed line), $m=0.5$ (dot-dashed line) and $m=1$ (continuous line) .} \label{Omega}
\end{figure}

We solve this equation numerically
and we choose different values for the parameter $\nu$ such that the difference between the roots  of the lapsus function $\Delta r=r_+-r_-$ decreases. Then, in Table \ref{QNM11}  we show the fundamental QNFs  for neutral massive scalar fields in the background of a hairy
black hole with $\kappa=0$, $\alpha_1=1$, $\alpha_2=2$, $k=1$ and different values of $m$ and $\nu$ and we plot these values in Fig. \ref{Omega}.
Furthermore,  we observe that in all the cases analyzed the QNFs have an imaginary part that is negative. So, we conclude that the radial massive neutral scalar field perturbations are stable in the
hairy black hole background and the system is always overdamped, that is, the QNFs have no real part so there is no oscillatory behaviour in the perturbations, only exponential decay.

 Also, for all cases analyzed, when we approach to the extremal limit $(\nu \approx 2.512)$, the absolute value of imaginary part decreases, see Fig. \ref{Omega}, which means that the damping time $\tau = 1/\omega_I$, where $\omega_I$ is the imaginary part of QNFs, is increasing. This result is interesting because it may have applications to the thermalization process of a quark-gluon plasma according to the gauge/gravity   duality. The recent theoretical and experimental
developments indicate  that the quark-gluon plasma produced at
Relativistic Hadron Ion Collider is a strongly interacting liquid
and the produced plasma locally isotropizes over a time scale of
$\tau_{iso} < \frac{1fm}{c} $. The dynamics of such
isotropization in a far-from-equilibrium non-Abelian plasma cannot
be described with the standard methods of field theory or
hydrodynamics. Then it was proposed in
\cite{Garfinkle:2011hm,Garfinkle:2011tc} that   such a
thermalization can be studied via its gravity dual identifying $\tau$ with the thermalization time.

     Besides, when the mass of the scalar field increases the absolute value of imaginary part increases and consequently the relaxation time decreases. It is worth mentioning, that in Table  \ref{QNM11},  we were able to compute the QNFs  until $\Delta r \approx 0.3$, for smaller values is difficult to find the QNFs by using the improved AIM. However, in the next section we will reach smaller values of $\Delta r$ for the near extremal black hole
and the extremal black hole by using the time domain analysis.

\begin{table}[ht]
\caption{Fundamental quasinormal frequencies for radial neutral scalar fields in the background of the charged hairy black hole with $\kappa=0$, $\alpha_1=1$, $\alpha_2=2$, $\Lambda=-0.1$, $q=0$, $k=1$ and different values of $m$ and $\nu$. }
\label{QNM11}\centering
\begin{tabular}{ | c | c | c | c | c | c | c |}
\hline
$\nu$ & $\Delta r$ & $m=0$ & $m=0.1$ & $m=0.5$ & $m=1$\\ \hline
$5.50$ & $7.48$ & $-1.29298 i$ &  $-1.31062 i$ & $-1.74474 i$ & $-2.61162 i$ \\
$5.00$ & $6.59$ & $-1.13103 i$ &  $-1.14816 i$ & $-1.53897 i$ & $-2.30644 i$ \\
$4.50$ & $5.66$ & $-0.96483 i$ &  $-0.98117 i$ & $-1.32610 i$ & $-1.99059 i$ \\
$4.00$ & $4.68$ & $-0.79231 i$ &  $-0.80735 i$ & $-1.10221 i$ & $-1.65809 i$ \\
$3.50$ & $3.61$ & $-0.60870 i$ &  $-0.62168 i$ & $-0.85916 i$ & $-1.29637 i$ \\
$3.00$ & $2.38$ & $-0.39963 i$ &  $-0.40917 i$ & $-0.57411 i$ & $-0.87034 i$ \\
$2.75$ & $1.60$ & $-0.26807 i$ &  $-0.27485 i$ & $-0.38946 i$ & $-0.59271 i$ \\
$2.70$ & $1.41$ & $-0.23609 i$ &  $-0.24213 i$ & $-0.34394 i$ & $-0.52401 i$ \\
$2.60$ & $0.95$ & $-0.15816 i$ &  $-0.16232 i$ & $-0.23207 i$ & $-0.35463 i$ \\
$2.55$ & $0.62$ & $-0.10248 i$ &  $-0.10524 i$ & $-0.15129 i$ & $-0.23179 i$ \\
$2.53$ & $0.42$ & $-0.06999 i$ &  $-0.07191 i$ & $-0.10380 i$ & $-0.15928 i$ \\
$2.525$ & $0.36$ & $-0.05935 i$ &  $-0.06099 i$ & $-0.08817 i$ & $-0.13537 i$ \\
$2.521$ & $0.30$ & $-0.04929 i$ &  $-0.05066 i$ & $-0.07336 i$ & $-0.11270 i$ \\ \hline
\end{tabular}%
\end{table}

\begin{figure}[h]
\begin{center}
\includegraphics[width=0.6\textwidth]{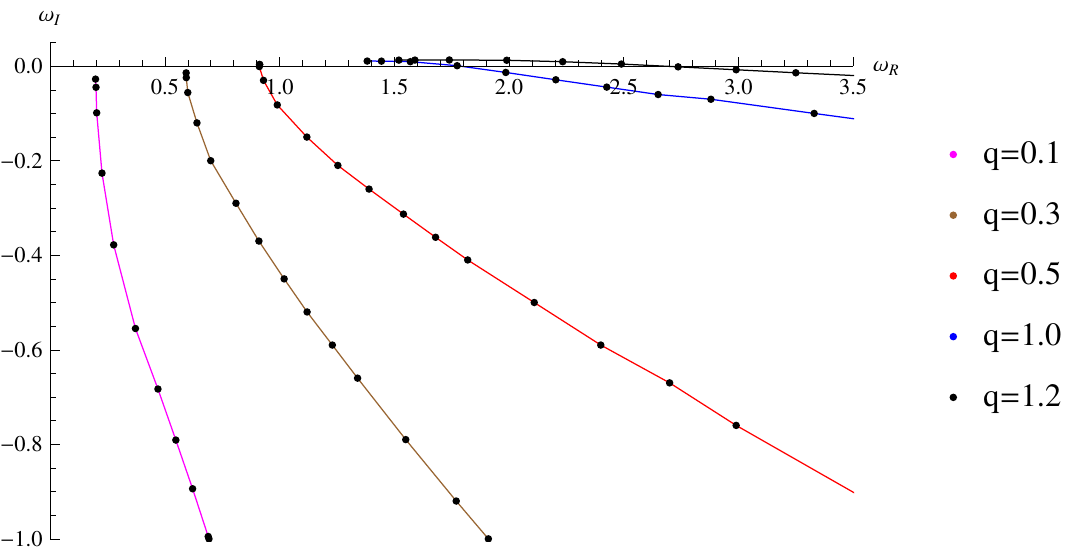}
\end{center}
\caption{The behaviour of $\omega$ in the complex plane for $\kappa=0$, $\alpha_1=1$, $\alpha_2=2$, $\Lambda=-0.1$, $m=0$, $
k=1$ and different values of $q$ and $\nu$.} \label{omegas}
\end{figure}

 On the other hand, for $\kappa=0$, $\alpha_1=1$, $\alpha_2=2$, $\Lambda=-0.1$, $m=0$, $k=1$ and  radial massless  charged scalar perturbations we give the  fundamental QNFs in Table \ref{QNM} and Table \ref{QNM1} for various values of $q$ and $\nu$ and we plot in the complex plane some of these values in Fig. \ref{omegas}. We observe that the QNFs present a real and imaginary part for massless charged scalar perturbations.  Also, for most of cases analyzed, when we approach to the extremal case, the value of real part decreases. Besides, we observe that as the charge $q$ is increasing more QNFs with positive imaginary part appear indicating an instability of the near extremal hairy black hole under radial massless charged perturbations. This behaviour can be seen clearly in Fig. \ref{omegas}.
In the next section we will show that these instabilities are due to the superradiant effect.
\begin{table}[ht]
\caption{Fundamental quasinormal frequencies for massless charged  scalar fields in the background of the  charged hairy black hole with $\kappa=0$, $\alpha_1=1$, $\alpha_2=2$, $\Lambda=-0.1$, $m=0$, $
k=1$ and different values of $q$ and $\nu$. }
\label{QNM}\centering
\begin{tabular}{ | c | c | c | c | c | c | c |}
\hline
$\nu$ & $\Delta r$ & $q=0$ & $q=0.1$ & $q=0.3$ & $q=0.5$ & $q=1$ \\ \hline
$30$ & $46.58$ & $-8.37759 i$ & $3.83217-5.73207 i$ & $6.55775-3.62443 i$ & $8.91096-2.38963 i$ & $14.02910-0.61098 i$ \\
$25$ & $38.75$ & $-6.96313 i$ & $3.19315-4.77281 i$ & $5.46656-3.01522 i$ & $7.42874-1.98619 i$ & $11.69600-0.50530 i$ \\
$20$ & $30.89$ & $-5.54380 i$ & $2.55402-3.81249 i$ & $4.37586-2.40459 i$ & $5.94733-1.58134 i$ & $9.36438-0.39858 i$ \\
$15$ & $23$ & $-4.11487 i$ & $1.91461-2.85004 i$ & $3.28614-1.79113 i$ & $4.46756-1.17364 i$ & $7.03560-0.28979 i$ \\
$10$ & $15.01$ & $-2.66288 i$ & $1.27414-1.88231 i$ & $2.19888-1.17044 i$ & $2.99201-0.75873 i$ & $4.71419-0.17594 i$ \\
$9$ & $3.51$ & $-2.36665 i$ & $1.14564-1.68736 i$ & $1.98209-1.04433 i$ & $2.69807-0.67380 i$ & $4.25195-0.15187 i$ \\
$8$ & $11.74$ & $-2.06700 i$ & $1.01676-1.49155 i$ & $1.76572-0.91697 i$ & $2.40489-0.58761 i$ & $3.79105-0.12701 i$ \\
$7$ & $10.07$ & $-1.76270 i$ & $0.88716-1.29452 i$ & $1.54996-0.78776 i$ & $2.11284-0.49962 i$ & $3.33213-0.10108 i$ \\
$6$ & $8.36$ & $-1.45183 i$ & $0.75601-1.09567 i$ & $1.33511-0.65569 i$ & $1.82259-0.40883 i$ & $2.87635-0.07367 i$ \\
$5.5$ & $7.48$ & $-1.29298 i$ &  $0.68917-0.99524 i$ & $1.22819-0.58802 i$ & $1.67851-0.36186 i$ & $2.65028-0.05926 i$ \\
$5$ & $6.59$ & $-1.13103 i$ &  $0.62056-0.89381 i$ & $1.12174-0.51875 i$ & $1.53548-0.31336 i$ & $2.42601-0.04431 i$ \\
$4.5$ & $5.66$ & $-0.96483 i$ &  $0.54830-0.79070 i$ & $1.01592-0.44719 i$ & $1.39400-0.26273 i$ & $2.20431-0.02886 $ \\
$4$ & $4.68$ & $-0.79231 i$ &  $0.46764-0.68287 i$ & $0.91091-0.37218 i$ & $1.25490-0.20897 i$ & $1.98637-0.01328 i$ \\
$3.5$ & $3.61$ & $-0.60870 i$ &  $0.37235-0.55477 i$ & $0.80689-0.29133 i$ & $1.11992-0.15023 i$ & $1.77398+0.00100 i$ \\
$3$ & $2.38$ & $-0.39963 i$ &  $0.27679-0.37773 i$ & $0.70326-0.19800 i$ & $0.99369-0.08226 i$ & $1.56765+0.00956 i$ \\
$2.7$ & $1.41$ & $-0.23609 i$ &  $0.22611-0.22627 i$ & $0.63818-0.12323 i$ & $0.92954-0.03180 i$ & $1.44390+0.01063 i$ \\
$2.55$ & $0.62$ & $-0.10248 i$ &  $0.20318-0.09867 i$ & $0.60083-0.05581 i$ & $0.91176-0.00084 i$ & $1.38163+0.01081 i$ \\
$2.52$ & $0.28$ & $-0.04645 i$ &  $0.19903-0.04470 i$ & $0.59345-0.02435 i$ & $0.91306+0.00338 i$ & $--$ \\
$2.515$ & $0.17$ & $-0.02848 i$ &  $0.19842-0.02738 i$ & $0.59285-0.01420 i$ & $0.91324+0.00355 i$ & $--$ \\ \hline

\end{tabular}%
\end{table}
\begin{table}[ht]
\caption{Fundamental quasinormal frequencies for massless charged scalar fields in the background of the charged hairy black hole with $\kappa=0$, $\alpha_1=1$, $\alpha_2=2$, $\Lambda=-0.1$, $m=0$, $
k=1$ and different values of $q$ and $\nu$. }
\label{QNM1}\centering
\begin{tabular}{ | c | c | c | c | c |}
\hline
$\nu$ & $q=1.2$ & $q=1.3$ & $q=1.5$ & $q=2$ \\ \hline
$30$ & $15.86600-0.23512 i$ & $16.73990-0.10562 i$ & $18.38480+0.05466 i$ & $21.87910+0.16806 i$ \\
$25$ & $13.22730-0.19309 i$ & $13.95560-0.08576 i$ & $15.32610+0.04671 i$ & $18.23650+0.14024 i$ \\
$20$ & $10.59010-0.15031 i$ & $11.17280-0.06530 i$ & $12.26880+0.03905 i$ & $14.59500+0.11248 i$ \\
$15$ & $7.95602-0.10605 i$ & $8.39322-0.04370 i$ & $9.21449+0.03192 i$ & $10.95550+0.08481 i$ \\
$10$ & $5.32988-0.05834 i$ & $5.62154-0.01952 i$ & $6.16737+0.02590 i$ & $7.32117+0.057382 i$ \\
$9$ & $4.80682-0.04796 i$ & $5.06934-0.01410 i$ & $5.55984+0.02490 i$ & $6.59566+0.05195 i$ \\
$8$ & $4.28515-0.03713 i$ & $4.51848-0.00838 i$ & $4.95345+0.02398 i$ & $5.87096+0.04656 i$ \\
$7$ & $3.76550-0.02574 i$ & $3.96956-0.00236 i$ & $4.34864+0.02309 i$ & $5.14741+0.04121 i$ \\
$6$ & $3.24896-0.01373 i$ & $3.42351+0.00391 i$ & $3.74608+0.02215 i$ & $4.42553+0.03592 i$ \\
$5.5$ & $2.99241-0.00756 i$ & $3.15203+0.00703 i$ & $3.44593+0.02158 i$ &  $4.06548+0.03330 i$ \\
$5$ & $2.73746-0.00139 i$ & $2.88190+0.00997 i$ &  $3.14675+0.02090 i$ & $3.70623+0.03072 i$ \\
$4.5$ & $2.48458+0.00447 i$ & $2.61342+0.01250 i$ &  $2.84872+0.02004 i$ & $3.34802+0.02817 i$ \\
$4$ & $2.23426+0.00943 i$ & $2.34681+0.01421 i$ & $2.55208+0.01898 i$ &  $2.99117+0.02567 i$ \\
$3.5$ & $1.98656+0.01245 i$ & $2.08202+0.01474 i$ & $2.25718+0.01775 i$ &  $2.63618+0.02322 i$ \\
$3$ & $1.74047+0.01307 i$ & $1.81894+0.01431 i$ & $1.96469+0.01648 i$ &  $2.28378+0.02085 i$ \\
$2.7$ & $1.59346+0.01292 i$ & $1.66226+0.01392 i$ & $1.79080+0.01572 i$ &  $2.07406+0.01945 i$ \\ \hline
\end{tabular}%
\end{table}

\begin{figure}[h]
\begin{center}
\includegraphics[width=0.5\textwidth]{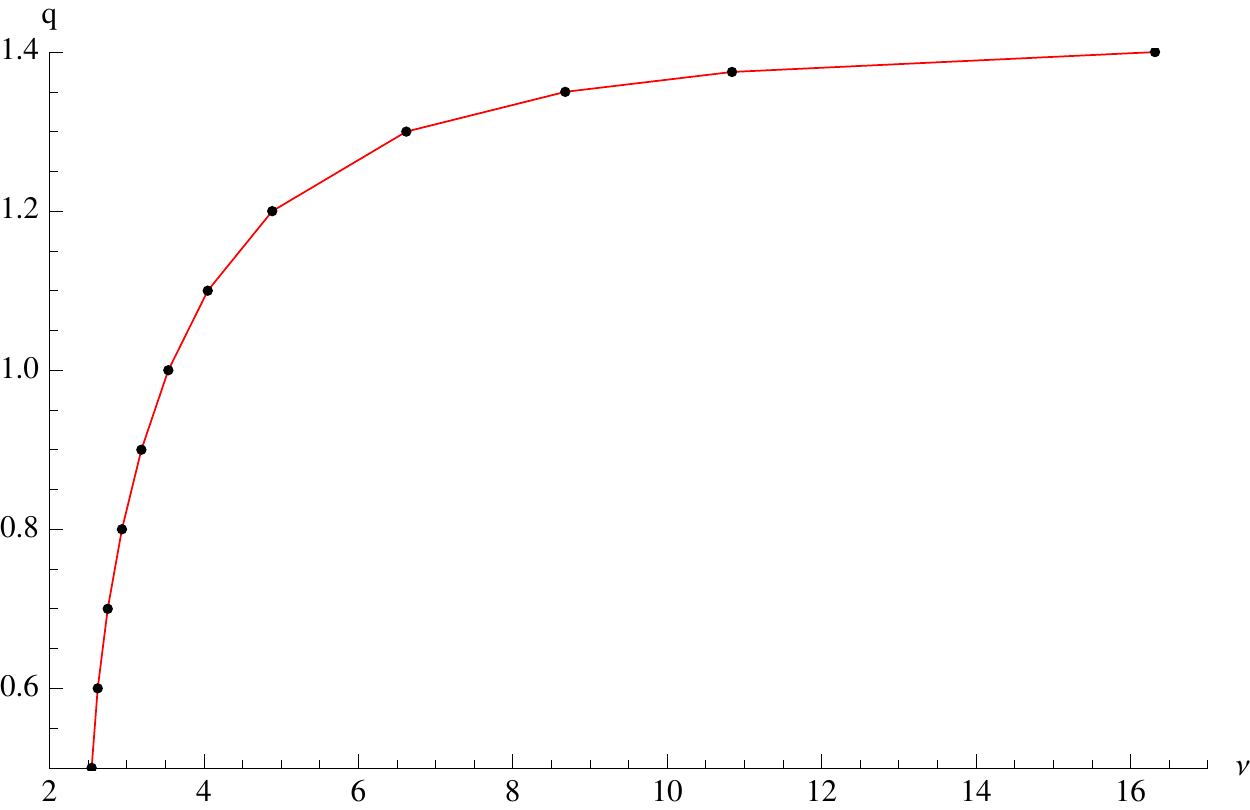}
\end{center}
\caption{Behaviour of $q_c$ (red curve) versus $\nu$ for $\kappa=0$, $\alpha_1=1$, $\alpha_2=2$, $\Lambda=-0.1$, $
k=1$ and $m=0$.  For $q>q_c$ the QNFs are unstable, for $q=q_c$ the QNFs have null imaginary part and for $q<q_c$ the QNFs are stable against radial massless charged scalar fields perturbations.} \label{sp}
\end{figure}

Furthermore, in Table \ref{QNM10} we obtain the relation between $q$ and $\nu$ for which the QNFs are purely real, for $\kappa=0$, $\alpha_1=1$, $\alpha_2=2$, $\Lambda=-0.1$, $m=0$, $
k=1$, and in Fig. \ref{sp} we depict  the values of Table \ref{QNM10}, i.e. $q$ as a function of $\nu$. We observe that for values of $q$ above the curve, the modes are unstable, while for values of $q$ below the curve, the modes are stable. Note that the values of $q$, $\nu$ and $\omega$ given in Table \ref{QNM10} satisfy the condition $\omega=q\sqrt{\alpha_1}\nu\log(1+\nu/r_+)$, which is closely related to the superradiance condition $\omega<q\sqrt{\alpha_1}\nu\log(1+\nu/r_+)$ that we will obtain in the next section,  and we will show that all the unstable modes are superradiant, while the stable modes are not superradiant. Additionally, the curve in Fig. \ref{sp} shows that for a fixed value of $q$ there is a limited value of $\nu$, such that above this limit the modes are stable and under this value the modes are unstable triggering the superradiance instability. Besides, for a fixed value of $\nu$ there is a maximum value of $q$ such that under this value the modes are stable and over this value the modes are unstable triggering the superradiance instability.
\begin{table}[h]
\caption{Fundamental quasinormal frequencies for massless charged  scalar fields with null imaginary part in the background of the charged hairy black hole with $\kappa=0$, $\alpha_1=1$, $\alpha_2=2$, $\Lambda=-0.1$, $m=0$, $k=1$ and different values of $q$. }
\label{QNM10}\centering
\begin{tabular}{ | c | c | c | c | c | c | c | c | c | c | c | c | c |}
\hline
$q$ & $0.5$ & $0.6$ & $0.7$ & $0.8$ & $0.9$ & $1.0$ & $1.1$ & $1.2$ & $1.3$ & $1.35$ & $1.375$ & $1.4$ \\ \hline
$\nu$ & $2.54565$ & $2.6248$ & $2.7544$ & $2.93848$ & $3.1902$ & $3.5392$ & $4.05$ & $4.88485$ & $6.6214$ & $8.6802$ & $10.8395$ & $16.318$ \\ \hline
$\omega$ & $0.91179$ & $1.04255$ & $1.18262$ & $1.34389$ & $1.53952$ & $1.79039$ & $2.13727$ & $2.67902$ & $3.76242$ & $5.01555$ & $6.31449$ & $9.58243$ \\ \hline
\end{tabular}%
\end{table}

\newpage

\subsubsection{Calculation of QNFs using the time domain analysis}
\label{TDA}

In this section we will apply a time evolution approach \cite{Wang:2000gsa} in order to compute the QNFs for the near extremal and the extremal hairy black hole, due to the failure of applicability of the improved AIM when we consider the extremal limit. It is worth mentioning that the QNFs for the extremal BTZ black hole cannot be obtained as limits of non-extremal BTZ black holes \cite{Crisostomo:2004hj}, and the imaginary part of the frequencies is zero (non-dissipative modes).  However, this behaviour can be explained because the change of variables in order to find an hypergeometric differential equation for the radial equation for the BTZ black hole, is not well defined when the two horizons coincide. In our case, the change of variables that we will perform is well defined for the extremal case. So, first it is convenient to express the metric in terms of the ingoing Eddington-Finkelstein coordinate $v=t+r^{\ast}$ \cite{Li:2015mqa, Li:2016kws}
\begin{equation}
ds^2=-f(r) dv^2+2 dv dr +a(r)^2 d \sigma^2~.
\end{equation}
The electromagnetic potential can be written as
\begin{equation}
A=\sqrt{\alpha_{1}} \nu \log \left( \frac{r}{r+ \nu} \right) dv~,
\end{equation}
and considering the following ansatz for the scalar field
\begin{equation}
\varphi=\frac{1}{a(r)} \phi (v,r) Y(\Omega)~,
\end{equation}
the Klein-Gordon equation \eqref{KGNM}, for a massless scalar field, yields
\begin{equation}
2 \partial_{v} \partial_{r} \phi +f(r) \partial_{r}^2 \phi+f^{\prime}(r) \partial_{r} \phi-2 i q A_{v} (r) \partial_{r} \phi - f^{\prime} (r) \frac{a^{\prime}(r)}{a(r)} \phi -i q A_{v} ^{\prime} (r) \phi-f(r) \frac{a^{ \prime \prime}(r)}{a(r)} \phi - \frac{\kappa^2}{a(r)^2} \phi = 0~.
\end{equation}
Performing the change of variable $z=\frac{2 r_{+}}{r}-1$, the above equation becomes
\begin{eqnarray}
\nonumber && \partial_{z} \partial_{v} \phi -\frac{1}{4 r_{+}} (1+z)^2 f(z) \partial_z^2 \phi-\left( \frac{1+z}{2 r_+} f(z) +\frac{(1+z)^2}{4 r_+} \frac{df}{dz} +i q A_{v}(z) \right) \partial_{z} \phi \\
&& -\left( -\frac{1}{4 r_+}(1+z)^2 \frac{df}{dz} \frac{1}{a}\frac{da}{dz}+ \frac{i q}{2} \frac{dA_{v}}{dz}-\frac{f}{a} \left( \frac{1+z}{2 r_+} \frac{da}{dz} +\frac{(1+z)^2}{4 r_+}  \frac{d^2 a}{dz^2} \right) -\frac{\kappa^2 r_+}{(1+z)^2 a^2}    \right) \phi=0~.
\end{eqnarray}

In the new coordinate the horizon is located at $z=1$ and the spatial infinity at $z=-1$. Now, in order to solve the above differential equation we use the pseudospectral Chebyshev method, see for instance \cite{Dias:2015nua,Yun}. The solution for the scalar field at any time is assumed to be a finite linear combination of the Chebyshev polynomials, with time dependent coefficients. Then, the solution is discretized at the Chebyshev collocation points, and the approximate solution is forced to satisfy the equation only at the Chebyshev collocation points. Futhermore, the exact derivatives are replaced by derivatives of interpolating polynomials at the Chebyshev points.

\begin{figure}[h]
\begin{center}
\includegraphics[width=0.4\textwidth]{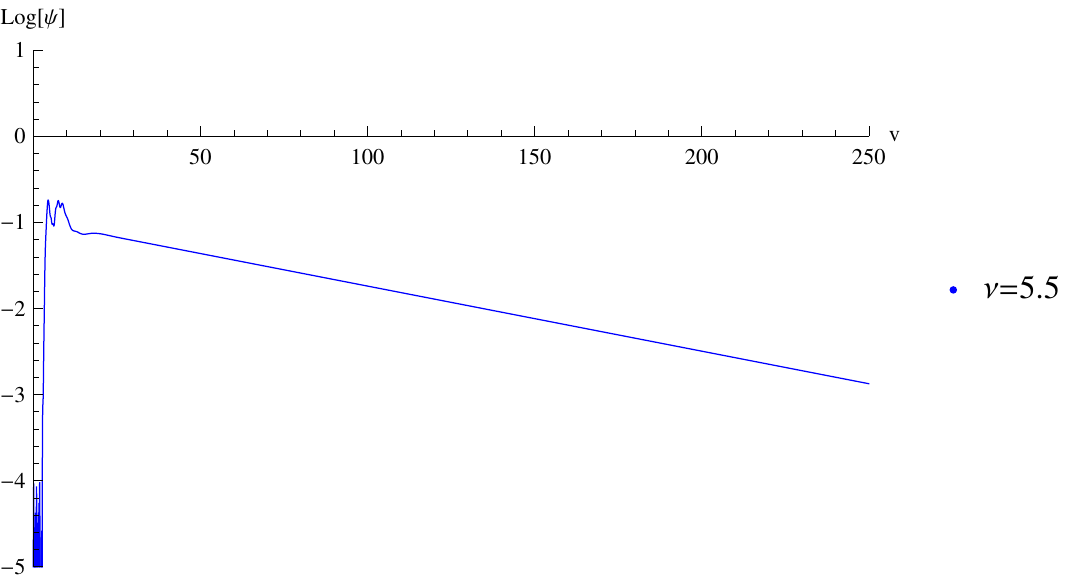}
\includegraphics[width=0.4\textwidth]{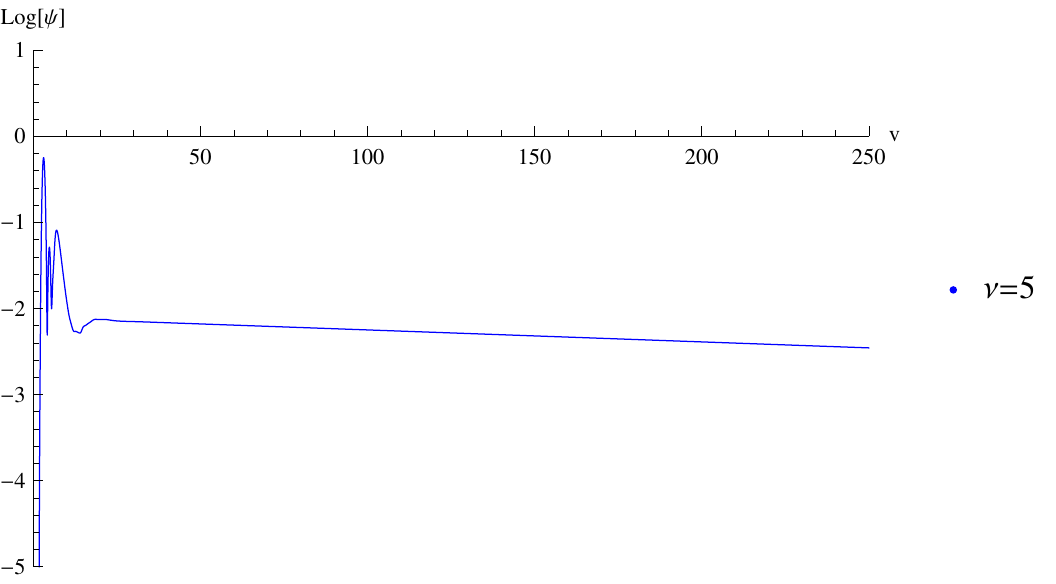}
\includegraphics[width=0.4\textwidth]{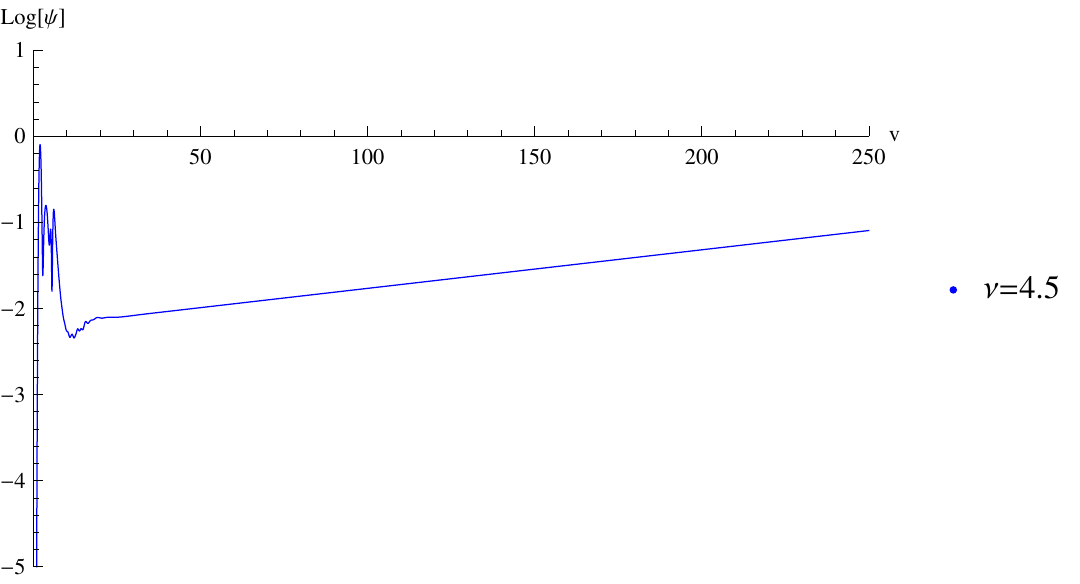}
\includegraphics[width=0.4\textwidth]{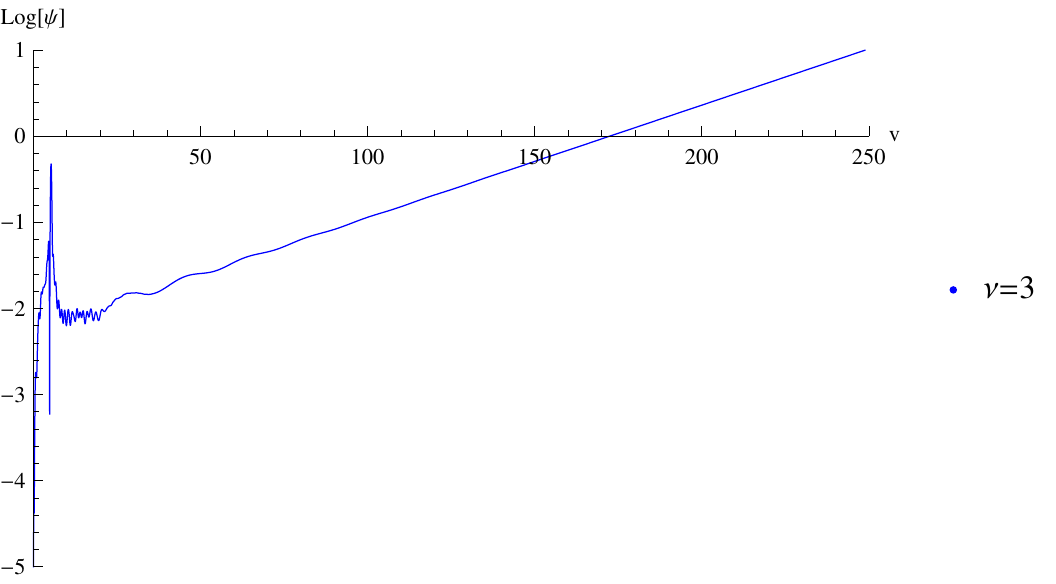}
\includegraphics[width=0.4\textwidth]{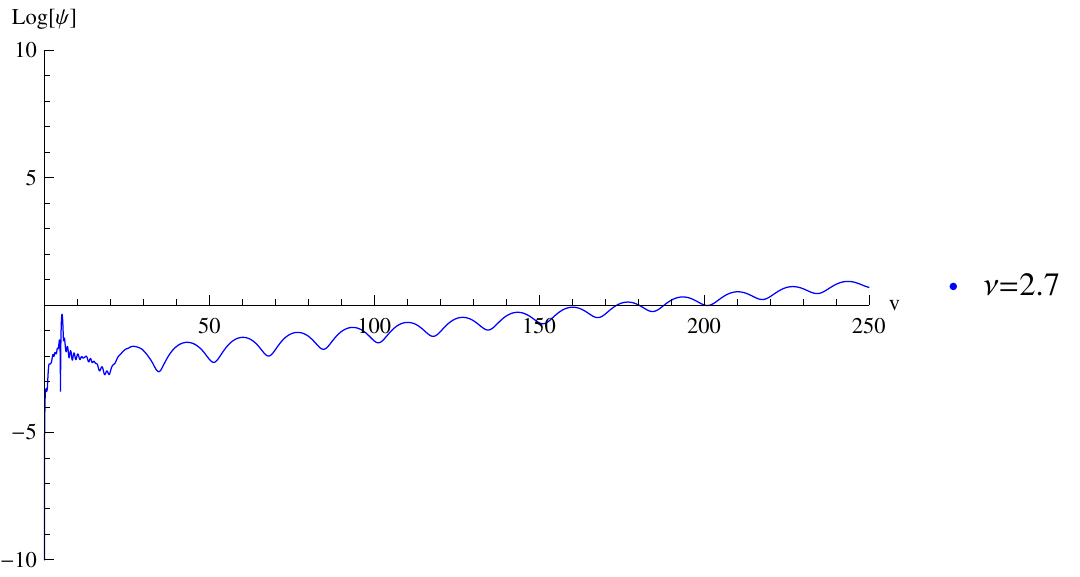}
\includegraphics[width=0.4\textwidth]{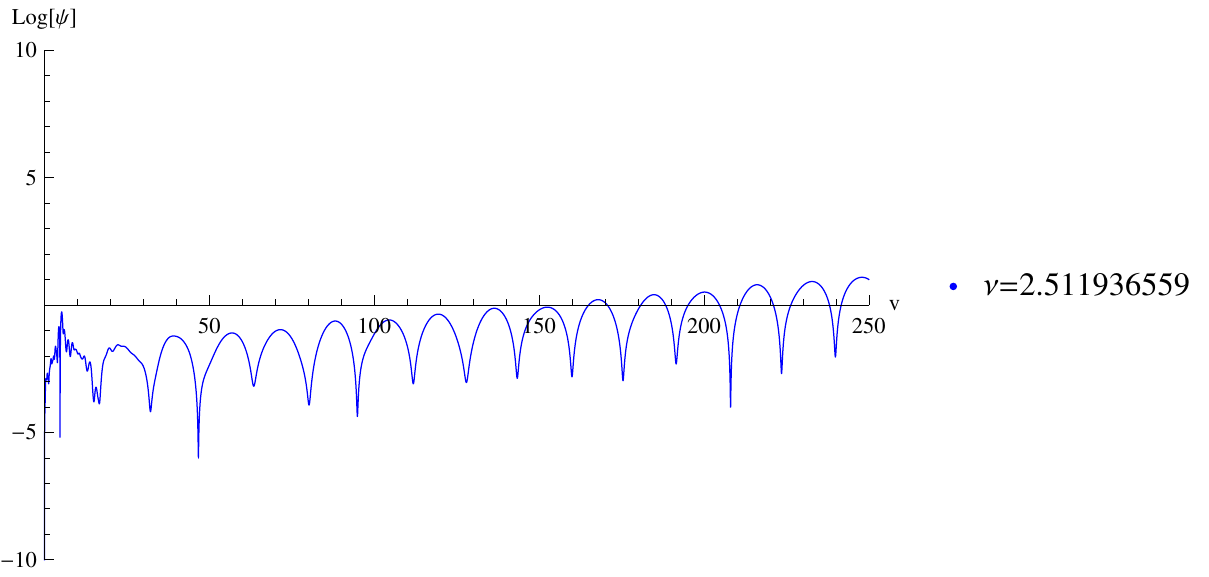}
\end{center}
\caption{Time evolution of charged massless scalar field for $\kappa=0$, $\alpha_1=1$, $\alpha_2=2$, $\Lambda=-0.1$, $q=1.2$, $
k=1$ and different values of $\nu$.} \label{profile}
\end{figure}

\begin{figure}[h]
\begin{center}
\includegraphics[width=0.4\textwidth]{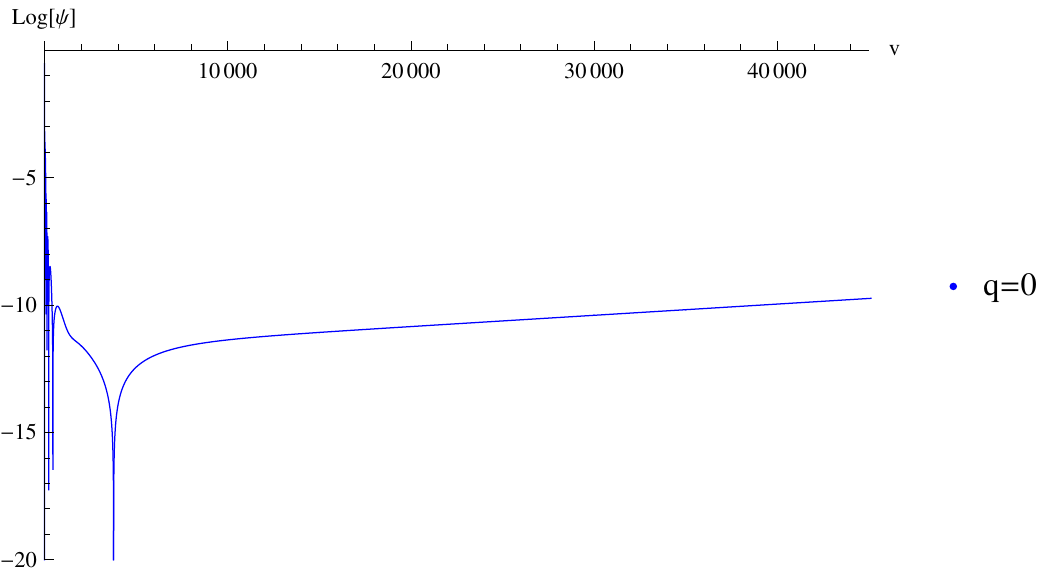}
\includegraphics[width=0.4\textwidth]{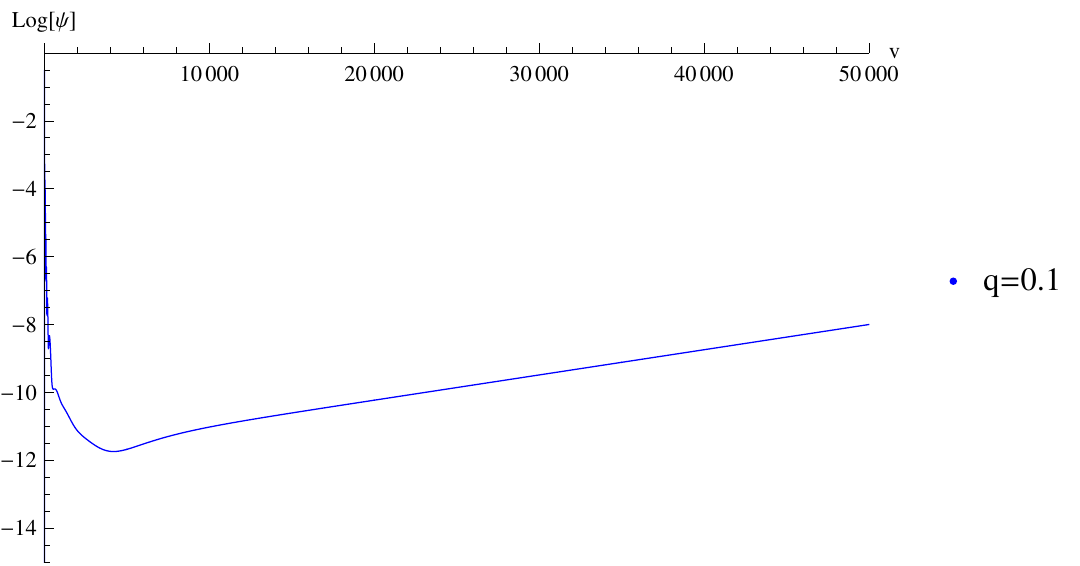}
\includegraphics[width=0.4\textwidth]{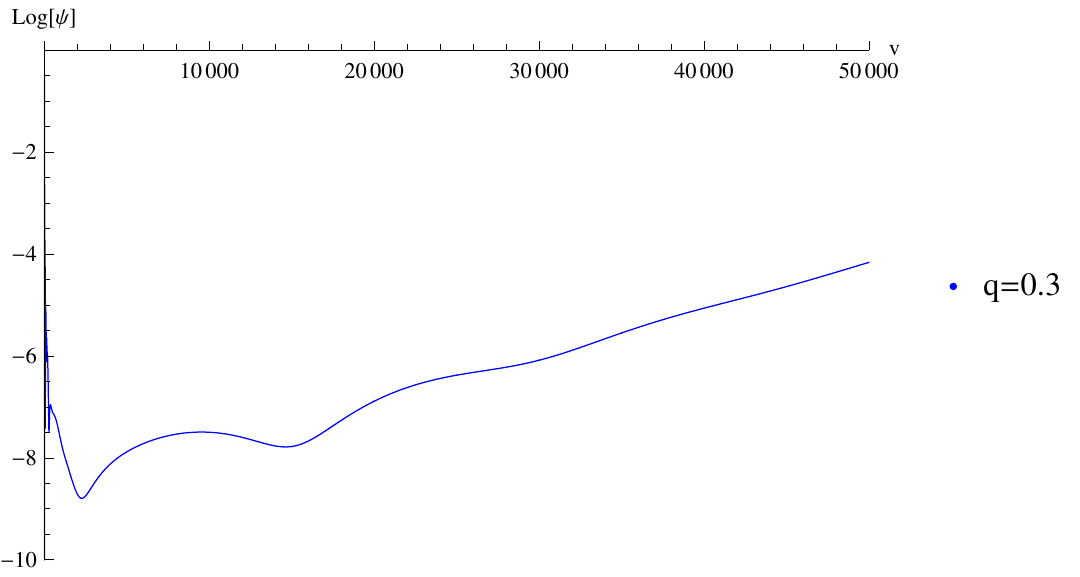}
\includegraphics[width=0.4\textwidth]{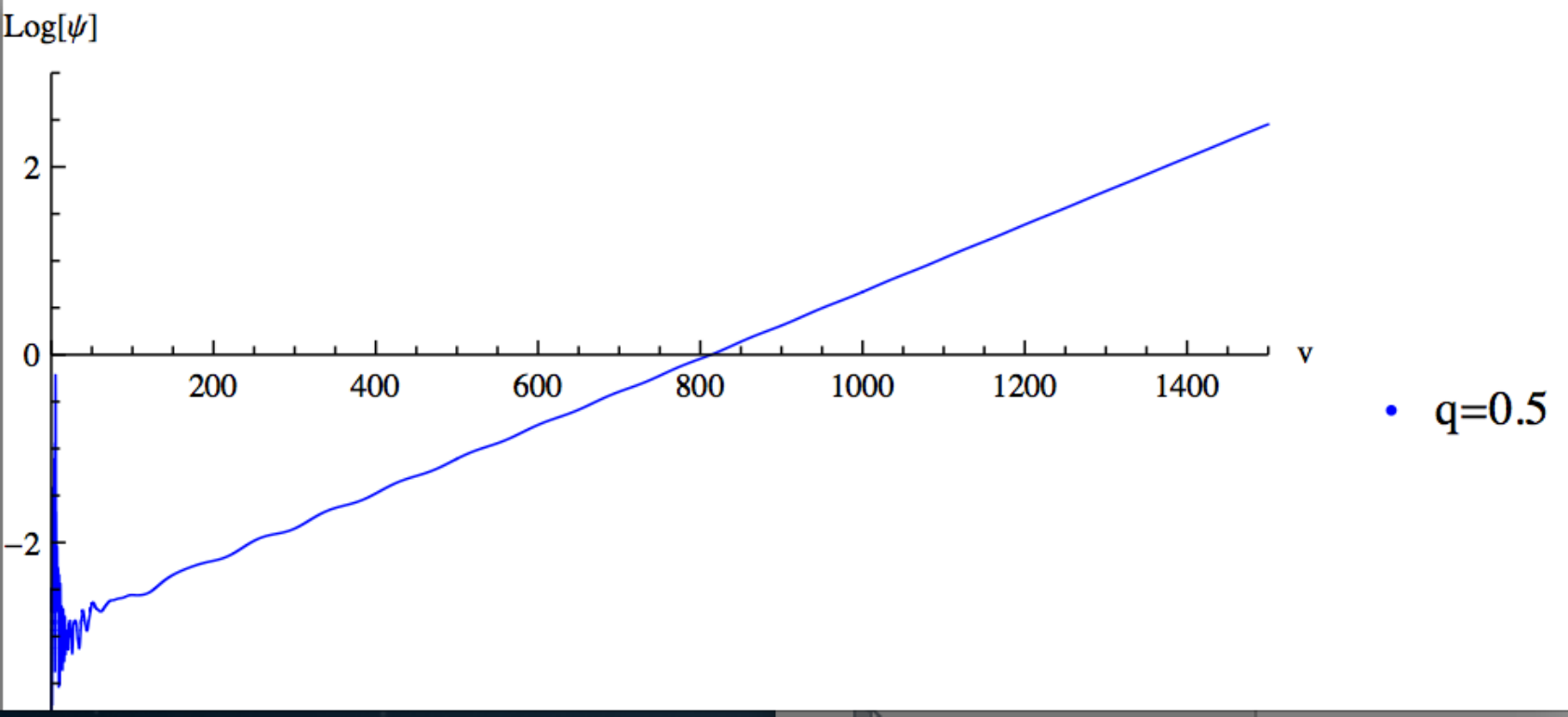}
\includegraphics[width=0.4\textwidth]{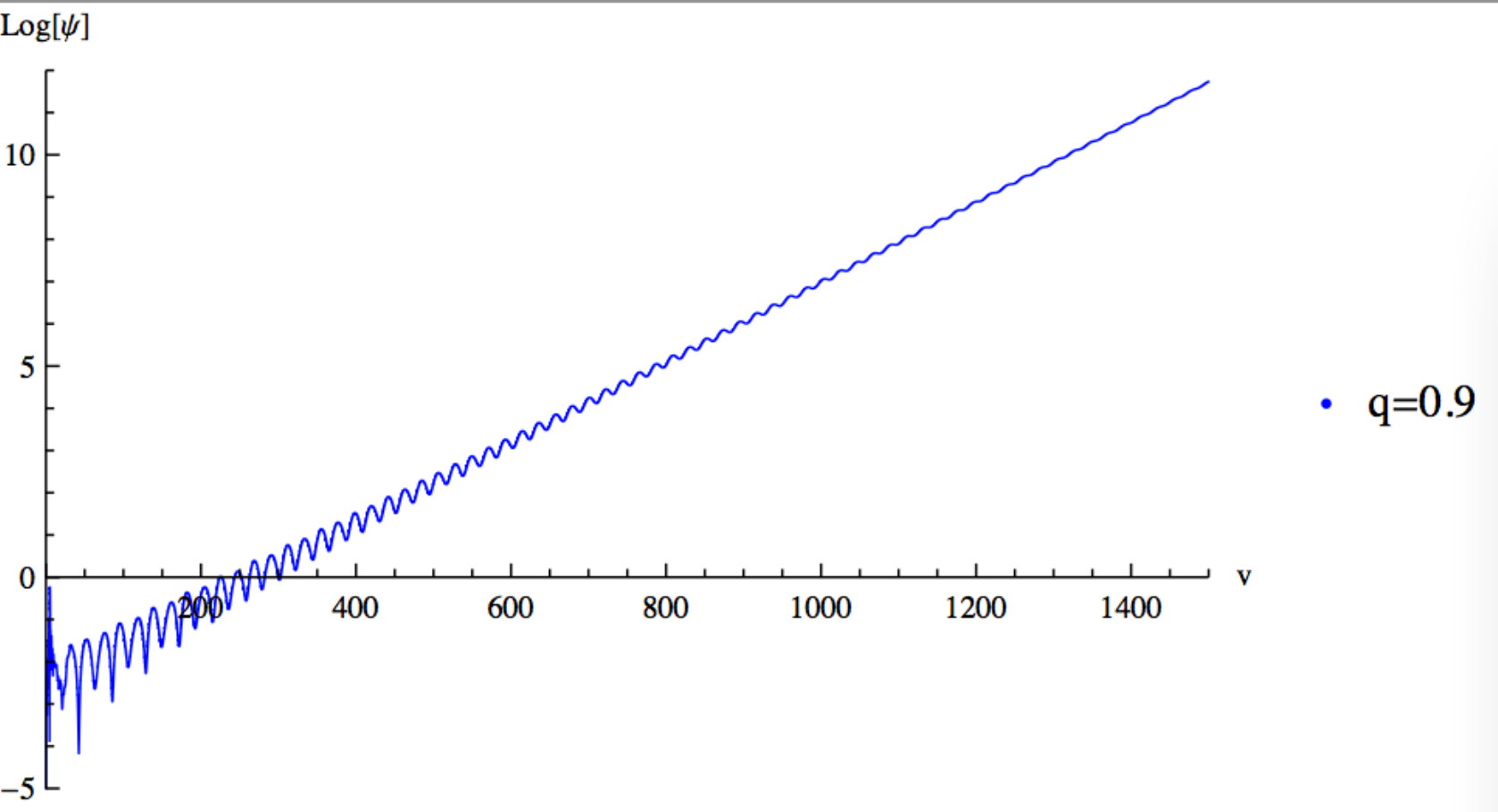}
\includegraphics[width=0.4\textwidth]{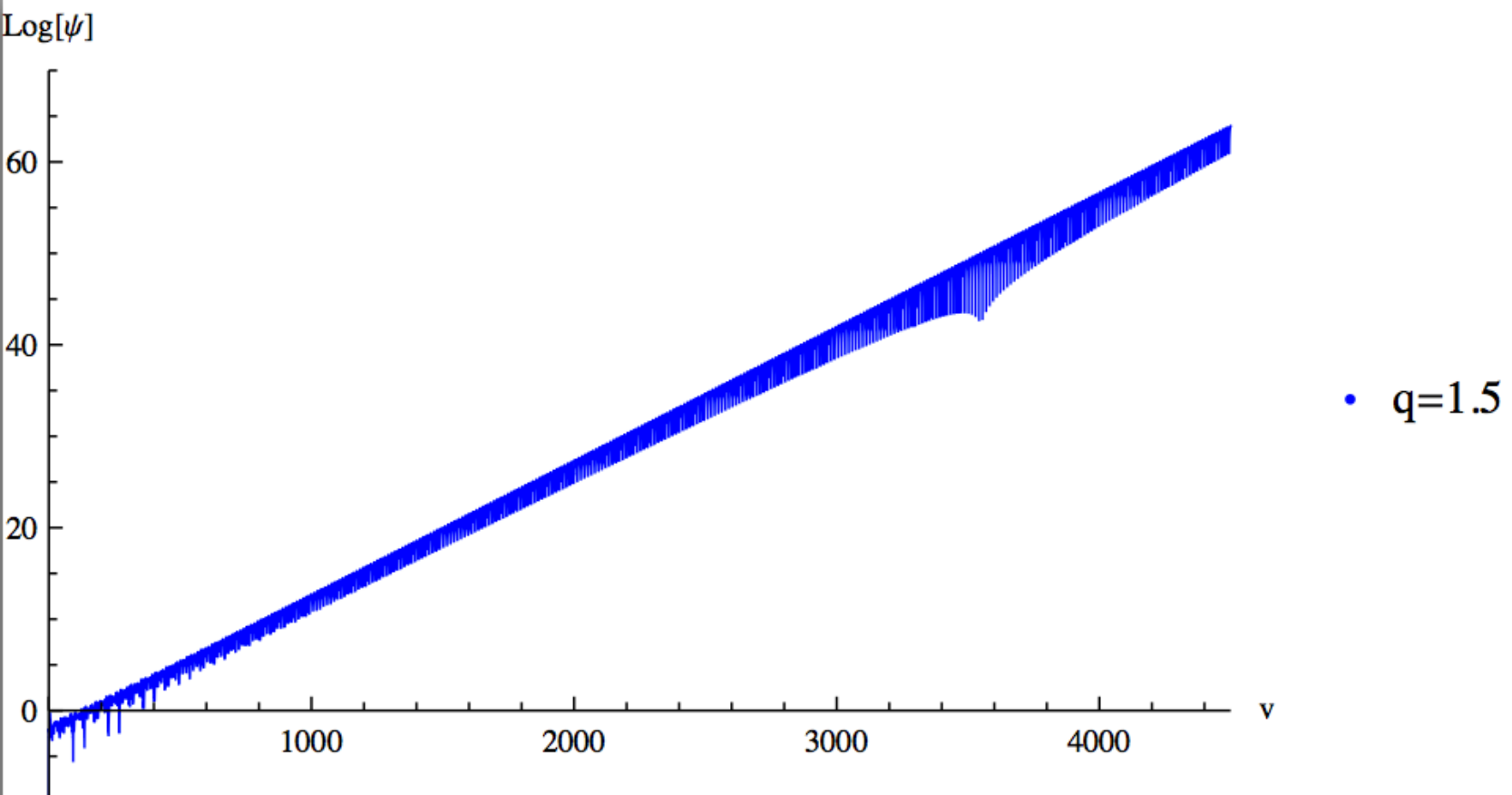}
\end{center}
\caption{Time evolution of charged massless scalar field for $\kappa=0$, $\alpha_1=1$, $\alpha_2=2$, $\Lambda=-0.1$, $\nu=2.511936559$ (Extremal case), $
k=1$ and different values of $q$.} \label{profile1}
\end{figure}

To integrate the system of differential equations in the time direction, we use the fourth order Runge-Kutta method. As initial condition for the scalar field we consider a Gaussian wave packet. Additionally, we set the boundary condition at spatial infinity as $\phi (v,z=-1)=0$. In Fig. \ref{profile} we show the evolution of massless scalar field in time,  we plot the logarithm of the absolute value of $\phi(v,r_0)$ vs $v$, for $r_0$ outside the horizon, and we can observe that the behaviour showed in the figures agrees with the results showed in Tables \ref{QNM} and \ref{QNM1}. In fact, we can see that for $\nu=5.5, 5$ the amplitude of the scalar field decreases in time which agrees with the results of the Tables because the imaginary part of the fundamental QNFs are negative in these cases. On the other hand, for $\nu=4.53, 2.7, 2.511936559$ the amplitude of the scalar field grows in time and the Tables show that the imaginary part of the QNFs are positive, which is due to a superradiant instability as we will show in the next section. Besides, In Fig. \ref{profile1}, we show the evolution of massless scalar field in time in the background of a extremal black hole, for $\kappa=0$, $\alpha_1=1$, $\alpha_2=2$, $\Lambda=-0.1$, $\nu=2.511936559$, $k=1$ and different values of $q$.  We can observe that the amplitude of the scalar field grows in time, although, for smaller values of the $q$, the scalar field grows more slowly than for large values of $q$, which also will be showed be due to a superradiant instability.

Now, we use the time profile data obtained from the time domain integration to calculate the predominant QNF using the Prony method, which fits the profile data using a superposition of damping exponents, \cite{Berti:2007dg, Zhidenko:2009zx, Li:2016kws}, in the following form

\begin{equation} \label{prony}
\phi (v,r) \approx \sum_{i=1}^p C_i e^{-i \omega_{i} (v-v_0)} \,.
\end{equation}

In the above equation, we have considered that the quasinormal ringing epoch starts at $v=v_0$ and ends at $v=v_0+Nh$, where $h$ denotes the step size in the time direction and the integer $N\ge 2p-1$. Formula \eqref{prony} is valid for each value from the profile data so, one obtain a set of equations

\begin{equation}
x_{n} \equiv \phi(nh+v_0,r)=\sum_{j=1}^p C_j e^{-i \omega_j nh}=\sum_{j=1}^p C_j z_j^n \,.
\end{equation}

The Prony method allows to find $z_i$ in terms of known $x_n$ and so, obtain the QNFs, which are given by the formula

\begin{equation}
w_j=\frac{i}{h} \ln (z_j) \,.
\end{equation}

\begin{figure}[h]
\begin{center}
\includegraphics[width=0.4\textwidth]{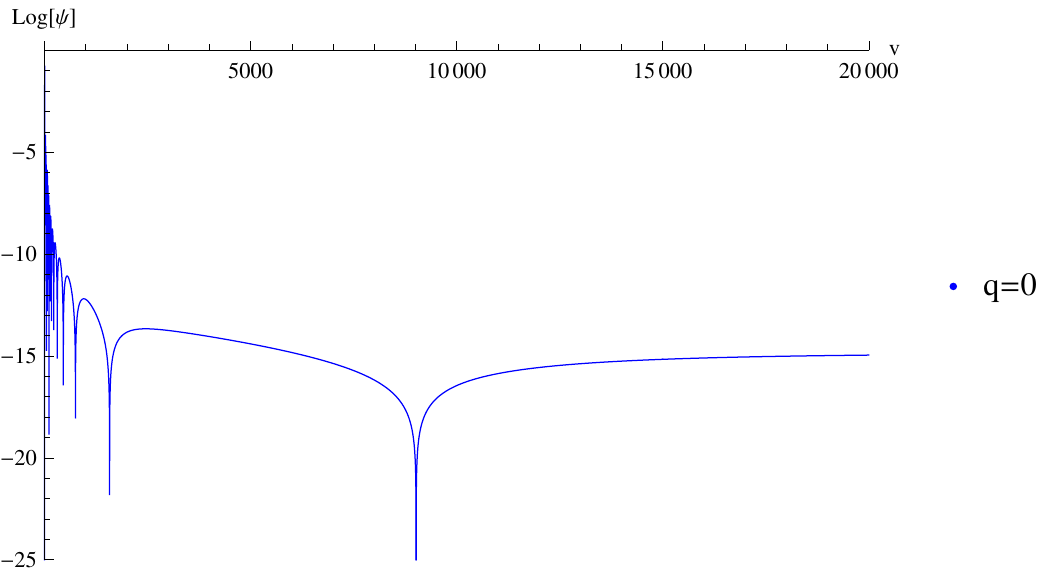}
\end{center}
\caption{Time evolution of charged massless scalar field for $\kappa=0$, $\alpha_1=1$, $\alpha_2=2$, $\Lambda=-0.1$, $\nu=2.511936559$ (Extremal case), $
k=1$, $q=0$ and 90 collocation points.} \label{grafico}
\end{figure}

In Table \ref{TD} we show some fundamental QNFs obtained by applying the Prony method to the time profile data, for $q=1.2$, $\kappa=0$, $\alpha_1=1$, $\alpha_2=2$, $\Lambda=-0.1$, $m=0$, $k=1$ and different values of $\nu$. We see that the values are in good agreement with the values obtained from the improved AIM. The QNFs coincide up to four decimals with respect to the improved AIM for the parameters considered. However with the time profile analysis we were able to obtain the QNFs for the near extremal and extremal black hole. Also,  in Table \ref{TD1} we have calculated QNFs for a nearly extremal black hole ($\nu=2.512$) and in Table \ref{TD2} for the extremal black hole  ($\nu=2.511936559$), for  $\kappa=0$, $\alpha_1=1$, $\alpha_2=2$, $\Lambda=-0.1$, $k=1$,  and different values of $q$. Note that for the nearly extremal black hole $\Delta r = 0.02$ and for the extremal black hole $\Delta r = 1.5 \cdot 10^{-8}$.  We observe that the scalar perturbations on the extremal black holes are always unstable, due to the imaginary part of the QNFs which is positive, as Table \ref{TD2} shows. For $q=0$ we see that the value of the QNF is approximately zero, and by increasing the number of collocation points from 50 to 90, we obtain  $\omega=4 \cdot 10^{-6}$ in this case which shows a convergence to zero as expected, see Fig. \ref{grafico}. Also, for $q=1$ and 90 collocation points we obtain $\omega=1.36584+0.01065 i$ which is in good agreement with the value shown in Table \ref{TD2}.
 It is worth to mention that the values of Table \ref{TD2}  agree with the observed in Fig. \ref{profile1}, that we have discussed previously.

\newpage

\begin{table}[ht]
\caption{Fundamental quasinormal frequencies for radial charged massless scalar fields in the background of the charged hairy black hole with $q=1.2$, $\kappa=0$, $\alpha_1=1$, $\alpha_2=2$, $\Lambda=-0.1$, $m=0$, $
k=1$ and different values of $\nu$.}
\label{TD}\centering
\begin{tabular}{ | c | c | c | c | c | c | c |}
\hline
$\nu$ & $\Delta r$ & $q=1.2$ \\ \hline
$5.50$ & $7.48$ & $2.99238-0.00756 i$ \\
$5.00$ & $6.59$ & $2.73744-0.00139 i$ \\
$4.50$ & $5.66$ & $2.48457+0.00448 i$ \\
$4.00$ & $4.68$ & $2.23426+0.00944 i$ \\
$3.50$ & $3.61$ & $1.98657+0.01246 i$ \\
$3.00$ & $2.38$ & $1.74045+0.01308 i$ \\
$2.70$ & $1.41$ & $1.59341+0.01295 i$ \\
$2.55$ & $0.62$ & $1.52013+0.01289 i$ \\
$2.511938$ & $0.0038$ & $1.50119+0.01232 i$ \\
$2.511936559$ & $1.5 \cdot 10^{-8}$ & $1.50114+0.01239 i$ \\ \hline
\end{tabular}%
\end{table}
\begin{table}[ht]
\caption{Fundamental quasinormal frequencies for charged massless scalar fields in the background of a nearly extremal charged hairy black hole with $\kappa=0$, $\alpha_1=1$, $\alpha_2=2$, $\Lambda=-0.1$, $k=1$, $\nu=2.512$ and different values of $q$. }
\label{TD1}\centering
\begin{tabular}{| c | c | c | c | c |}
\hline
$q=0.3$ & $q=0.35$ & $q=0.5$ & $q=0.9$ & $q=1$ \\ \hline
$0.59376-0.00135 i$ & $0.68937+0.00029 i$ & $0.91333+0.00363 i$ & $1.29178+0.00987 i$ & $1.36537+0.01067 i$ \\ \hline
\end{tabular}%
\end{table}
\begin{table}[ht]
\caption{Fundamental quasinormal frequencies for charged massless scalar fields in the background of an extremal  charged hairy black hole with $\kappa=0$, $\alpha_1=1$, $\alpha_2=2$, $\Lambda=-0.1$, $k=1$, $\nu=2.511936559$ and different values of $q$. }
\label{TD2}\centering
\begin{tabular}{ | c | c | c | c | c | c | c | c |}
\hline
$q=0$ & $q=0.3$ & $q=0.35$ & $q=0.4$  & $q=0.45$ \\ \hline
$0.00004 i$ & $0.59423+0.00009 i$ &  $0.68973+0.00036 i$ & $0.77461+0.00134 i$  &  $0.84821+0.00257 i$ \\ \hline
$q=0.5$ & $q=0.9$ & $q=1$ & $q=1.2$ & $q=1.5$ \\ \hline
$0.91326+0.00357 i$ &  $1.29171+0.00948 i$ & $1.36536+0.01113 i$ & $1.50114+0.01239 i$ & $1.68163+0.01470 i$ \\ \hline
\end{tabular}%
\end{table}

\begin{figure}[h]
\begin{center}
\includegraphics[width=0.6\textwidth]{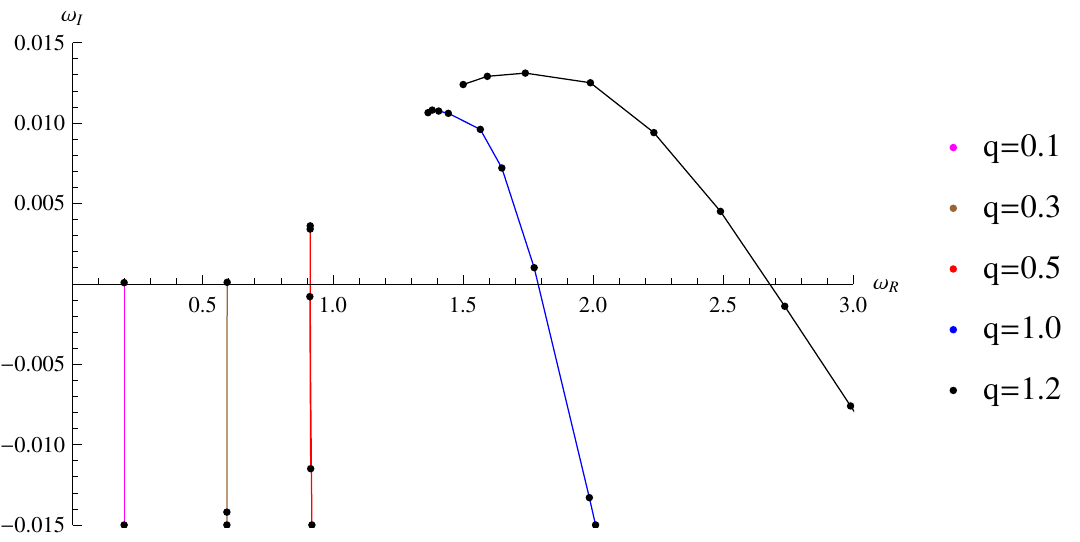}
\end{center}
\caption{The behavior of $\omega$ in the complex plane for $\kappa=0$, $\alpha_1=1$, $\alpha_2=2$, $\Lambda=-0.1$, $m=0$, $
k=1$ and different values of $q$ and $\nu$.} \label{Final}
\end{figure}

Now having the QNFs from the time domain analysis we can improve the behaviour of the Fig. \ref{omegas} that we have showed in the improved AIM section. So, in Fig. \ref{Final} we plot the behavior of $\omega$ in the complex plane, by considering the transition from a hairy black hole to the extremal hairy black hole, and we can observe as the QNFs converge to the QNFs for the extremal case and that for some cases, when the QNFs have a positive imaginary part, the imaginary part present a maximum value and then decay, while that in other cases, for small values of the charge $q$,  the imaginary part of the QNFs always increases.

\newpage

\section{Superradiant effect}
\label{superrandiant}

In this Section we will study the superrandiant effect of a radial massive charged scalar field of the form (\ref{wave}) scattered off the horizon of the charged hairy black hole (\ref{redfun}). In a pioneering work  Bekenstein showed \cite{Bekenstein}
that if the condition
\be  \omega<q\Phi~, \label{bek} \ee
is satisfied, where $q$ is the charge coupling constant of the field and
$\Phi$ is the electric  potential of the charged black hole,  then the superradiant scattering
of charged scalar field  results in the
extraction of Coulomb energy and electric charge from the charged
black hole. Then, this amplification of
charged massive fields by charged black holes  leads to an
instability of the Reissner-Nordstr\"om spacetime.

This charged radiation extracted from the  event horizon of the black hole has to traverse a nontrivial,
curved spacetime geometry before it eventually reaches asymptotic infinity.
This radiation procedure if it takes place in an AdS space  works
as a potential barrier for the radiation, giving a deviation from the blackbody radiation spectrum and this deviation is measure by the greybody factor. To be able to measure this  energy we have to trap these scattered
waves in a cavity which usually is created by an effective potential outside the black hole horizon and then from the absorption
and scattering cross sections of these waves compute their corresponding coefficients (for a recent review on superradiance see \cite{Brito:2015oca}).

To find the conditions for superradiance ampplification of scatter wave we will compute the greybody factor and the reflection coefficients. Then  if the greybody factor is negative or the reflection coefficients is greater than 1 \cite{Benone:2015bst} then the scalar waves can be superradiantly amplified by black hole. Following  \cite{Harmark:2007jy} we will split the spacetime in three regions and we will consider the low frequency limit, that is $\omega+qA_t(r_+) < < T_H$ and $(\omega+qA_t(r_+)) r_+ << 1$, and by simplicity we consider $\ell=0$.
\begin{itemize}
\item
Region I: The region near the event horizon, defined by $r \approx r_+$ and $V_{eff}(r) \approx - 2q \omega A_t(r_+)-q^2 A_t(r_+)^2$ or $V_{eff}(r)|_{q=0} << (\omega+qA_t)^2$.
In the first region, the solution to the radial equation is given by equation (\ref{eq1}), that can be rewritten as
\begin{equation}
R(r)=A_Ie^{-i(\omega+qA_t(r_+))r^*}~,
\end{equation}
which slightly away the horizon yields
\begin{equation}
R(r)=A_I \left(1-\frac{i(\omega+qA_t(r_+))}{f'(r_+)}\log \left( \frac{r-r_+}{r_+}\right) \right)~.
\end{equation}

\item
Region II: The intermediate region, between the horizon region and the asymptotic region. This region is defined by $V_{eff}(r)|_{q=0}>>(\omega+qA_t)^2$.\\

In this case the radial equation \eqref{radial} reduces to
\begin{equation}
\frac{d}{dr}\left(a(r)^2f(r)\frac{dR}{dr}\right)=0~,
\end{equation}
whose solution is given by
\begin{equation}
R(r)=A_{II}+B_{II}G(r)~,
\end{equation}
where
\begin{equation}
G(r)=\int_{\infty}^r \frac{dr}{f(r)a(r)^2}~.
\end{equation}
So, for $r\approx r_+$ we obtain
\begin{equation}
G(r)\approx\frac{1}{f'(r_+)r_+(r_++\nu)}\log\left(\frac{r-r_+}{r_+}\right)~.
\end{equation}
Matching this solution with the solution of region I, we obtain
\begin{equation}
A_{II}=A_{I},\,\,\, B_{II}=-i(\omega+qA_t(r_+))r_+(r_++\nu)A_{I}~.
\end{equation}

On the other hand, for $r>>r_+$
\begin{equation}
G(r)\approx\int_{\infty}^r \frac{dr}{-\frac{\Lambda r^2}{3}a(r)^2}=-\frac{1 }{\Lambda r^3}.
\end{equation}
Therefore
\begin{equation}
R(r)=A_I\left(1-\frac{i(\omega+qA_t(r_+))r_+ (r_++\nu)}{\Lambda r^3}\right),
\end{equation}
which will be matched with asymptotic behaviour.
\item
Region III: The asymptotic region. This region is defined by $r>>r_+$.\\

In this case, for our purposes it is enough to take into account just the leading term in the asymptotic behaviour of the effective potential, which is given by
\begin{equation}
V_{eff}(r)\approx  \frac{2 \Lambda^2 r^2}{9}~,
\end{equation}
and the tortoise coordinate at infinity is given by
\begin{equation}
r^*=\frac{2}{\sqrt{4+\Lambda \nu^2/3}\sqrt{-\Lambda/3}}\arctan \left(\frac{(2r+\nu)\sqrt{-\Lambda/3}}{\sqrt{4+\Lambda \nu^2/3}} \right)~.
\end{equation}
In this way, the solution of the radial equation in the asymptotic region is
\begin{equation}
R(r)= C_1+C_2/r^3~.
\end{equation}
Then, matching the solution of region II, for $r>>r_+$, with the solution of region III yields
\begin{equation}
C_1=A_I, \,\,\,  C_2= -i(\omega+qA_t(r_+))r_+(r_++\nu)A_I/\Lambda~.
\end{equation}
\end{itemize}
Now, we will compute the fluxes, that are given by the following expression
\begin{equation}
\mathcal{F}=\frac{\sqrt{-g}g^{rr}}{2i}\left( R^*\partial_r R-R\partial_r R^*\right)~.
\end{equation}
So, at the horizon we have
\begin{equation}
\mathcal{F}_{hor}\propto -(\omega+qA_t(r_+))r_+ (r_++\nu)|A_I|^2~.
\end{equation}
and at infinity by
\begin{equation}
\mathcal{F}_{\infty}\propto -\Lambda Im ( C_1C_2^* )~.
\end{equation}
In order to characterize the fluxes  at the asymptotic region, is convenient to split up the coefficients $C_{1}$ and $C_{2}$ in terms of the incoming and outgoing coefficients, $\hat{C}_2$ and $\hat{C}_{1}$, respectively. We define $C_1=\hat{C}_1+\hat{C}_2$ and $C_2=i(\hat{C}_2-\hat{C}_1)$. Therefore, the asymptotic  incoming and outgoing fluxes are given by

\begin{equation}
\mathcal{F}_{in\,\, \infty}\propto \Lambda |\hat{C}_2|^2=\Lambda \frac{|A_I|^2}{4}\left(1-\frac{(\omega+qA_t(r_+))r_+(r_++\nu)}{\Lambda}\right)^2~,
\end{equation}

\begin{equation}
\mathcal{F}_{out\,\, \infty}\propto -\Lambda |\hat{C}_1|^2=-\Lambda \frac{|A_I|^2}{4}\left(1+\frac{(\omega+qA_t(r_+))r_+(r_++\nu)}{\Lambda}\right)^2~.
\end{equation}
Then, the reflection coefficient and the greybody factor \cite{Birmingham:1997rj,  Kim:1999un, Harmark:2007jy, Oh:2008tc, Kao:2009fh, Gonzalez:2010ht, Gonzalez:2010vv, Gonzalez:2011du} yields
\begin{equation}
\mathcal{R}= \frac{\mathcal{F}_{out\,\, \infty}}{\mathcal{F}_{in\,\, \infty}}=\left(\frac{1+\frac{(\omega+qA_t(r_+))r_+(r_++\nu)}{\Lambda}}{1-\frac{(\omega+qA_t(r_+))r_+(r_++\nu)}{\Lambda}} \right)^2~,
\end{equation}
\begin{equation}
\gamma(\omega)= \frac{\mathcal{F}_{hor}}{\mathcal{F}_{in\,\, \infty}}=\frac{\frac{-4(\omega+qA_t(r_+))r_+(r_++\nu)}{\Lambda}}{\left(1-\frac{(\omega+qA_t(r_+))r_+(r_++\nu)}{\Lambda}\right)^2}~.
\end{equation}
Finally, it is possible to find the superradiant condition from the reflection coefficient or from the greybody factor. So the reflection coefficient is greater than 1 or alternatively the greybody factor is negative if $\omega+qA_t(r_+)<0$ or
\be \omega<q\sqrt{\alpha_1}\nu\log(1+\nu/r_+) \label{nbek}~.\ee Note that this condition because the parameter of the scalar field $\nu$ is related to the charge $Q$ of the black hole  (relation (\ref{redef})), modifies the Bekenstein's superradiance condition (\ref{bek}) introducing the information of the   presence of the scalar hair.  The analysis was performed for the non-extremal case; however, the same condition (\ref{nbek}) still holds for the extremal black hole.
Then the comparison of  the real part of the dominant modes, that we have found in the previous section, with the superradiant condition of charged scalar fields \cite{Cardoso:2004hs, Cardoso:2006wa, Aliev:2008yk, Dias:2011at, Li:2012rx, Uchikata:2011zz, Cardoso:2013pza} suggests
 that all the unstable modes are superradiant unstable and consequently the scalar waves can be superradiantly amplified by the black hole.

\section{Conclusions}
\label{conclusion}

We have studied the QNFs of radial scalar field perturbations of near extremal and extremal four-dimensional charged hairy black holes in AdS spacetime. First, we have studied the case of hairy black hole by using the improved AIM and we have found the fundamental QNFs for radial neutral scalar field perturbations. In this case, for all cases analyzed, we have showed that the system is overdamped, that is the QNFs have no real part so there is no oscillatory behaviour in the perturbations, only exponential decay. Moreover, the imaginary part of the QNFs is negative, which guarantees the stability of the propagation of massive neutral scalar fields in such background. However, when the distance between the two roots of the lapsus function decreases, it is more difficult to find the QNFs by using the improved AIM. So, in order to obtain the QNFs for the near extremal and extremal black hole, we have applied the time domain analysis. In general, for the cases analyzed, we have showed that the QNFs of the near extremal black hole converge to the QNFs for the extremal black hole.

 When we considered the propagation of massless charged scalar fields we have found that the QNFs present real and imaginary parts, even in some cases the imaginary part is positive, which indicates that the propagation of charged massless scalar fields is unstable. Then, we have studied the superradiance effect  calculating the superradiant condition. We have found that the superrandiat condition depends on $\nu$, the charge of the scalar field that backreacts on the geometry, giving hair to the  charged black hole. The presence of $\nu$ in the superradiant condition, which is proportional to $Q$ the charge of the hairy black hole, modifies in a non-trivial way the Bekenstein's superradiance condition. We have also found that the critical value of the charged scalar perturbations $q_c$ giving unstable modes, is related to the charge of the scalar hair $\nu$ and there is a maximum value of $\nu$ which can trigger the supperrandiant instability.
Finally, we have found that all the unstable modes are superradiant  and all the stable modes are not superradiant, satisfying the superradiant condition, and consequently the radial scalar waves can be superradiantly amplified by black hole by extracting charged of the black hole which indicates that the black hole is unstable.

  It is worth mentioning that it would be interesting to analyze the superradiant instability of this hairy black hole for charged massive scalar field with non vanishing angular momentum, due to the fact that in some cases the instability is sensitive to the parity of the orbital quantum number (for instance see \cite{Aliev:2008yk}), which we left for a future work.

\acknowledgments

We thank Emanuelle Berti and Bin Wang for their useful comments and remarks.  This work was partially funded by Comisi\'{o}n
Nacional de Ciencias y Tecnolog\'{i}a through FONDECYT Grants 11140674 (PAG) and by the Direcci\'{o}n de Investigaci\'{o}n y Desarrollo de la Universidad de La Serena (Y.V.).
P. A. G. acknowledges the hospitality of the Universidad de La Serena, National Technical University of Athens and Pontificia Universidad Cat\'{o}lica de Valpara\'{i}so,
E. P. acknowledges the hospitality of the Universidad Diego Portales and Pontificia Universidad Cat\'{o}lica de Valpara\'{i}so and Y. V. acknowledges the hospitality of the Pontificia Universidad Cat\'{o}lica de Valpara\'{i}so  where part of this work was undertaken.

\appendix

\section{Improved AIM}
In this appendix we give a brief review of the improved AIM, which is used to solve
homogeneous
linear second-order differential equations subject to boundary conditions. First, it is necessary to implement the boundary conditions. For this purpose the dependent variable must be redefined in terms of a new function, say $\chi$, that satisfies the boundary conditions appropriate to the eigenvalue problem under consideration. Thus, in order to implement the improved AIM the differential equation must be written in the form
\begin{equation}
\chi^{\prime \prime }=\lambda _{0}(y)\chi ^{\prime }+s_{0}(y)\chi \,.
\label{de}
\end{equation}%
Then,
one must
to differentiate Eq. (\ref{de}) $n$ times with respect to $y$,
which yields the following equation:
\begin{equation}
\chi^{n+2}=\lambda _{n}(y)\chi ^{\prime }+s_{n}(y)\chi~,  \label{de1}
\end{equation}%
where
\begin{equation}
\lambda _{n}(y)=\lambda _{n-1}^{\prime }(y)+s_{n-1}(y)+\lambda
_{0}(y)\lambda _{n-1}(y)~,  \label{Ln}
\end{equation}%
\begin{equation}
s_{n}(y)=s_{n-1}^{\prime }(y)+s_{0}(y)\lambda _{n-1}(y)\,.  \label{Snn}
\end{equation}%
Then, expanding the $\lambda _{n}$ and $s_{n}$ in a Taylor series around
some point $\eta $, at which the improved AIM is performed, yields
\begin{equation}
\lambda _{n}(\eta )=\sum_{i=0}^{\infty }c_{n}^{i}(y-\eta )^{i}\,,
\end{equation}%
\begin{equation}
s_{n}(\eta )=\sum_{i=0}^{\infty }d_{n}^{i}(y-\eta )^{i}\,,
\end{equation}%
where the $c_{n}^{i}$ and $d_{n}^{i}$ are the $i^{th}$ Taylor coefficients
of $\lambda _{n}(\eta )$ and $s_{n}(\eta )$, respectively, and by replacing
the above expansions in Eqs. (\ref{Ln}) and (\ref{Snn}) the following
set of recursion relations for the coefficients is obtained
\begin{equation}
c_{n}^{i}=(i+1)c_{n-1}^{i+1}+d_{n-1}^{i}+%
\sum_{k=0}^{i}c_{0}^{k}c_{n-1}^{i-k}~,
\end{equation}%
\begin{equation}
d_{n}^{i}=(i+1)d_{n-1}^{i+1}+\sum_{k=0}^{i}d_{0}^{k}c_{n-1}^{i-k}\,.
\end{equation}%
In this manner, the authors of the improved AIM have avoided the
derivatives that contain the AIM in \cite{Cho:2009cj, Cho:2011sf}, and
the quantization condition, which is equivalent to imposing a termination
to the number of iterations, is given by
\begin{equation}
d_{n}^{0}c_{n-1}^{0}-d_{n-1}^{0}c_{n}^{0}=0~.
\end{equation}



\end{document}